\documentclass{article}

\PassOptionsToPackage{numbers, compress}{natbib}
\usepackage[preprint]{neurips_2024}

\usepackage[utf8]{inputenc} % allow utf-8 input
\usepackage[T1]{fontenc}    % use 8-bit T1 fonts
\usepackage{hyperref}       % hyperlinks
\usepackage{url}            % simple URL typesetting
\usepackage{booktabs}       % professional-quality tables
\usepackage{amsfonts}       % blackboard math symbols
\usepackage{nicefrac}       % compact symbols for 1/2, etc.
\usepackage{microtype}      % microtypography
\usepackage{xcolor}         % colors

\usepackage{amsmath}
\usepackage{wrapfig}
\usepackage{graphicx}
\usepackage{bbm}
\usepackage{algorithm}
\usepackage{algorithmic}

\usepackage{graphicx}
\usepackage{natbib}
\usepackage{doi}

\usepackage{microtype}
\usepackage{graphicx}
\usepackage{subfigure}
\usepackage{booktabs} % for professional tables
\usepackage{multicol}
\usepackage{bm}
\usepackage{hyperref}
\usepackage{enumitem}
\usepackage{diagbox}
\usepackage{amsmath}
\usepackage{amssymb}
\usepackage{mathtools}
\usepackage{amsthm}
\usepackage{hyperref}
\usepackage{url}
\usepackage{graphicx}
\usepackage{multirow}
\usepackage{tcolorbox}
\tcbuselibrary{breakable}
\usepackage[toc]{appendix}
\usepackage{minitoc}
\usepackage[capitalize,noabbrev]{cleveref}
\usepackage{authblk}
\usepackage{enumitem}
\usepackage{float}

\usepackage{amsmath}
\DeclareMathOperator*{\argmax}{arg\,max}

\usepackage{subfigure}
\usepackage{parskip}
\usepackage{xcolor}

\usepackage{float}
\title{Safe Multi-agent Reinforcement Learning with Natural Language Constraints}
\author{
\textbf{Ziyan Wang}$^{1}$, 
\textbf{Meng Fang}$^{2}$, 
\textbf{Tristan Tomilin}$^{3}$, 
\textbf{Fei Fang}$^{4}$,
\textbf{Yali Du}$^{1}$ \\\
$^1$ King's College London 
$^2$ University of Liverpool \\
$^3$ Eindhoven University of Technology 
$^4$ Carnegie Mellon University \\\
\vspace{0.0cm}
\texttt{ziyan.wang@kcl.ac.uk, meng.fang@liverpool.ac.uk,} \\
\texttt{t.tomilin@tue.nl, feifang@cmu.edu, yali.du@kcl.ac.uk}
}

\begin{document}

\maketitle

\begin{abstract}
The role of natural language constraints in Safe Multi-agent Reinforcement Learning (MARL) is crucial, yet often overlooked. While Safe MARL has vast potential, especially in fields like robotics and autonomous vehicles, its full potential is limited by the need to define constraints in pre-designed mathematical terms, which requires extensive domain expertise and reinforcement learning knowledge, hindering its broader adoption. To address this limitation and make Safe MARL more accessible and adaptable, we propose a novel approach named \textbf{S}afe \textbf{M}ulti-\textbf{A}gent Reinforcement \textbf{L}earning with natural \textbf{L}anguage constraints (\textbf{SMALL}). Our method leverages fine-tuned language models to interpret and process free-form textual constraints, converting them into semantic embeddings that capture the essence of prohibited states and behaviours. These embeddings are then integrated into the multi-agent policy learning process, enabling agents to learn policies that minimize constraint violations while optimizing rewards. To evaluate the effectiveness of SMALL, we introduce the LaMaSafe, a multi-task benchmark designed to assess the performance of multiple agents in adhering to natural language constraints. Empirical evaluations across various environments demonstrate that SMALL achieves comparable rewards and significantly fewer constraint violations, highlighting its effectiveness in understanding and enforcing natural language constraints.

% The potential of Safe Multi-agent Reinforcement Learning is vast, especially in fields like robotics and autonomous vehicles. However, its full potential is limited by the need to define constraints in mathematical terms, which requires both domain expertise and extensive RL knowledge and limits its broader adoption. To address this limitation, we proposed a novel approach named \textbf{S}afe \textbf{M}ulti-\textbf{A}gent \textbf{R}einforcement \textbf{L}earning with natural \textbf{L}anguage constraints (\textbf{SMALL}), which interprets natural language descriptions of constraints. Our approach converts textual constraints into representations of prohibited states, enabling the learning of a policy aimed at minimizing the violations of these constraints during training. To test the effectiveness of our approach, we introduced the LaMaSafetyGoal, a multi-task benchmark designed to evaluate the performance of multiple agents tasked with optimizing rewards while adhering to constraints specified in free-form text. Empirical evaluations across various environments in LaMaSafetyGoal demonstrate that our method outperforms existing approaches by achieving higher rewards and fewer constraint violations. Extensive ablation studies further highlight the crucial role of fine-tuned language models in improving both coordination and safety in multi-agent systems. 
\end{abstract}

% \begin{figure}[t]
%     \centering
%     \includegraphics[width=1.\linewidth, trim=0 7 0 7, clip]{figs/figure.pdf}
%     \vspace{-0.2in}
%     \caption{Learning curves on a suite of MuJoCo benchmark tasks with episodic rewards, based on 5 independent runs with random initialization. The shaded region indicates the standard deviation and the curves were smoothed by averaging the 10 most recent evaluation points using an exponential moving average. An evaluation point was established every $10^4$ time steps.}
% \label{fig:main_results}
% \end{figure}

% \begin{wrapfigure}{r}{0.65\textwidth}
%     \vspace{-0.05in}
%     \centering
%     \includegraphics[width=1.\linewidth]{figs/visualization-rewards.png}
%     \vspace{-0.25in}
%     \caption{The visualization of decomposed rewards (blue solid lines) and the grounded rewards (red dotted lines).}
%     \vspace{-0.2in}
%     \label{fig:visualized_rewards}
% \end{wrapfigure}

% \begin{wrapfigure}{r}{0.45\textwidth}
%     \centering
%     \vspace{-0.1in}
% \includegraphics[width=0.97\linewidth]{figs/causal_graph.pdf}
%     \vspace{-0.08in}
%     \caption{A graphical illustration of causal structure in Generative Return Decomposition. See the main text for the interpretation.}
%     \vspace{-0.1in}
% \label{fig:graphical_env_model}
% \end{wrapfigure}

\section{Introduction}

\begin{figure*}[ht] \centering\includegraphics[width=0.95\textwidth]{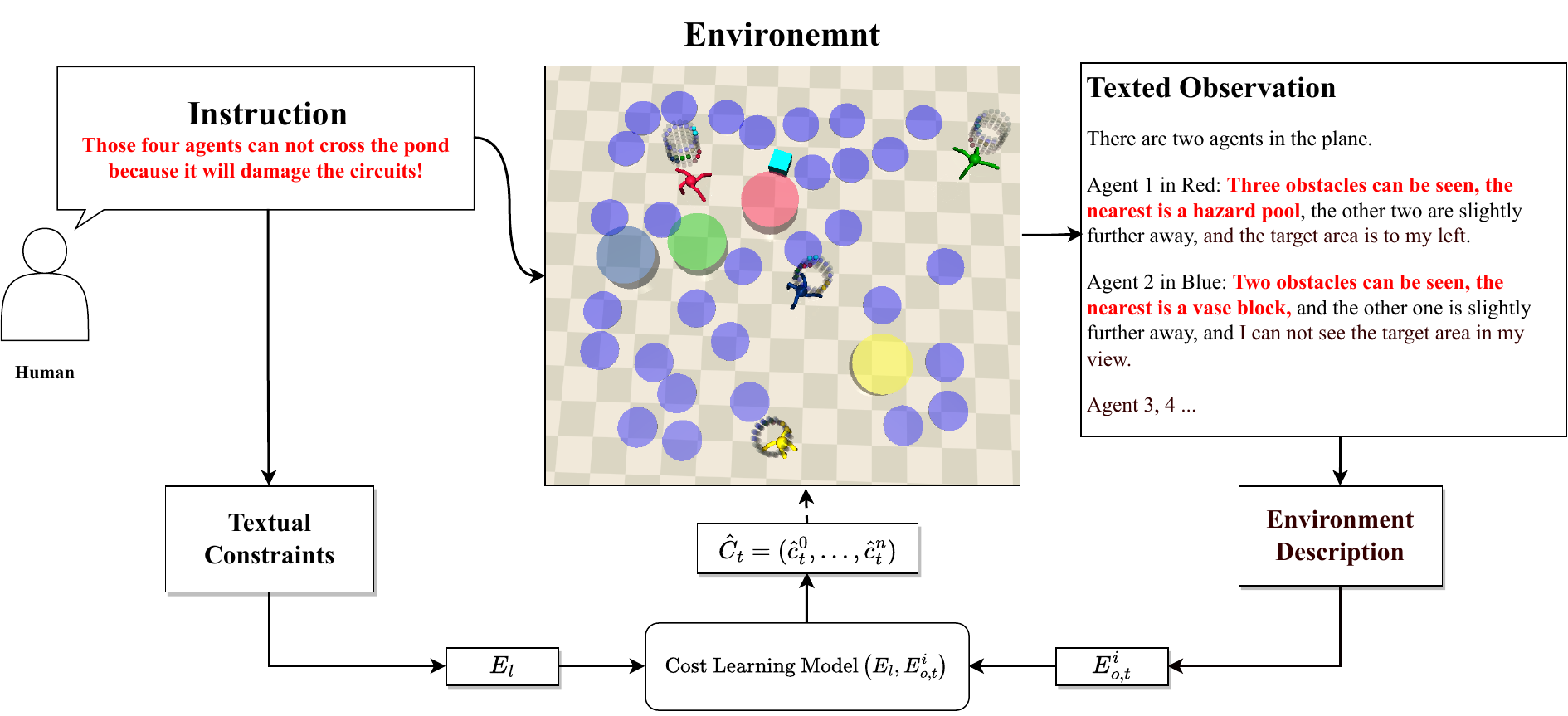}
    %\caption{Structure of the proposed method. An autoregressive LLM (decoder only LLM) and an encoder-only LLM is adopted to model the constraint transformation function $M$. $\widehat{M}$ extracts semantic meaning of constraint $x$ and outputs the parameter to approximate cost function $\hat{C}$.}
    %\caption{Cost prediction in the proposed method. 
    \caption{ The framework of the \textbf{SMALL}. Initially, humans will create natural language constraints for the environment and agents. Firstly, SMALL uses the decoder language model to condense the semantic meaning of the nature of human instruction and eliminate ambiguity and redundancy. Secondly, the encoder Language Model encodes the condensed constraints and environment description from the text-based observations into embeddings $E_l$ and ${E}^i_{o,t}$ according to their semantic meaning. Lastly, the cost prediction model uses those embeddings as input and predicts the constraint is violated (predicted the cost $\hat{c}^n_t$ for each agent). In the end, the policy network will update using the prediction cost and the embeddings.}
    \label{fig:struc}
\vspace{-0.5cm}
\end{figure*}

% \mf{say multi-agent RL: Recent years have witnessed significant advancements in multi-agent reinforcement learning (MARL) across various domains xxx } 

% Recent years have witnessed significant advancements in multi-agent reinforcement learning (MARL) across domains such as robotics control~\cite{}, resource allocation~\cite{} and achieving expert level in games~\cite{}. In real-world settings such as  and , MARL agents should not be given complete freedom for safety, fairness or other concerns and must adhere to specific constraints provided by humans. To this end, safe RL follows the constrained Markov Decision Process (CMDP) \cite{altman1999constrained} framework, where agents must satisfy the constraints given by auxiliary costs when optimizing their policies to maximize the objective function. 

%%% Start
In recent years, Multi-agent Reinforcement Learning (MARL) has shown great potential in various challenging problems such as robotics control~\citep{perrusquia2021multi,peng2021facmac} and mastering complex games~\citep{vinyals2019grandmaster, du2019liir}. In practical scenarios such as resource balancing~\citep{li2019cooperative}, traffic management~\citep{waymax} and healthcare systems~\citep{shaik2023ai}, MARL agents must operate within strict boundaries due to safety, fairness, or ethical considerations. A surge in interest in safe MARL has led to the rise of learning algorithms that optimize agents' policies for maximum efficacy while adhering to human-imposed constraints. However, current safe MARL approaches rather only consider a fixed format barrier or factored shielding function generated by the prior knowledge ~\citep{cai2021safe,elsayed2021safe} or only consider the settling pre-designed cost function~\citep{liu2021cmix,gu2021multi,lu2021decentralized}.

The role of natural language constraints in Safe MARL is crucial, but it is often ignored. Human languages offer an intuitive and easily accessible medium for describing constraints, not only for machine learning experts or system developers but also for potential end-users who interact with agents such as household robots. However, a shortcoming of current safe MARL methods lies in their inability to adapt to the nuances of human language constraints. First, the natural language constraints are challenging to estimate and incorporate into a numerical cost function due to their diverse and context-specific nature.In many real-world scenarios, unlike the designed cost functions, humans often implement a range of natural language constraints to address security concerns preemptively. This is particularly true if new language-based constraints emerge, tailored to specific needs and situations, which these pre-designed cost functions fail to anticipate. For example, a language constraint could be stated as, `Avoid blue obstacles, which indicate danger.' By articulating constraints in natural language, users can easily define safety standards, operational limits, and ethical boundaries, making the technology more accessible and controllable.  Second, the challenge of adapting to linguistic constraints is considerably magnified due to the complexity of inter-agent decision-making, where multiple agents need not only to understand and respond to language-based instructions individually but also maintain cooperation with other agents. The inability of pre-designed cost functions to adapt to these rapidly evolving and diverse language constraints can lead to significant operational inefficiencies and increased risks. 
%Furthermore, determining whether all agents or the specific agents are subject to these natural language constraints without a specific cost function becomes unclear. 

%To learn the safety multi-agent reinforcement learning policy via human language, prior related methods~\cite{yang2021safe} in the single agent setting have attempted to learn relevant entities from the environment to predict constraint violations, but this can lead to computational inefficiency and inaccuracies in complex settings. Furthermore, some methods fix the structure of the constraints ~\cite{prakash2020guiding} and train them in a supervised learning manner but make them ineffective under novel and free-form natural language input. \zy{Check, those are all single agent methods; should we mention them or cite them?}
% \ff{I would suggest you move the first two sentences in this paragraph to the previous paragraph (before you talk about the technical challenges).}

%To this end, to demonstrate learning under this setting, 
In this paper, we propose a new method for learning a policy with language constraint prediction to address the challenge of safe MARL with natural language, named \textbf{S}afe \textbf{M}ulti-\textbf{A}gent \textbf{R}einforcement \textbf{L}earning with natural \textbf{L}anguage constraints (\textbf{SMALL}). As illustrated in Figure \ref{fig:struc}, initially, we employ the large language model (LLM) to summarize the linguistic description of the constraints, aiming to align them with the environment setting and disambiguate them to extract semantics. LLMs~\citep{brown2020language, touvron2023llama} are fine-tuned on extensive corpora and are consistent with human values, making them adept at extracting high-quality constraints. Subsequently, we utilize a cost learning module to learn how well the natural language constraints align with the textual descriptions of the environment and to estimate the cost of constraint violations based on semantic similarities. This allows agents to adjust their policies to enforce these constraints while learning the task.
Additionally, to evaluate the effectiveness of our approach, we have developed the first safe multi-agent benchmark incorporating natural language constraints, termed LaMaSafe.

To summarize, our paper presents three main contributions. First, we are the first to introduce safe MARL with natural language constraints. This significantly improves the traditional cost function approach by allowing complex, free-form constraints for safer and more adaptable multi-agent scenarios. Second, we have developed LaMaSafe, a pioneering benchmark for more realistic safety constraint scenarios in MARL. This benchmark is designed to rigorously evaluate the performance of various algorithms under the unique challenges posed by free-form natural language constraints. Third, we introduce SMALL, a novel method for enhancing safety in MARL environments. The empirical results in both discrete and continuous action settings environments demonstrate that SMALL can achieve comparable rewards to other MARL algorithms while significantly reducing constraint violations.

\section{Related Work}
% \yali{add a summary paragraph describing what you are going to introduce next and the motivation why introducing them.}

In this section, we will explore three interconnected areas: the relatively safe MARL and its baselines, the role of natural language in enhancing MARL to follow human instructions, and the work on Language Models related to our approach.

\textbf{Safe Multi-Agent Reinforcement Learning: } Despite the significant attention given to safe MARL in recent years, many safety-related challenges remain unresolved~\citep{gu2022review}, such as dealing with natural language constraints. Several approaches have been proposed to address the safety problem by using fixed pre-design cost functions. The safe model-free MARL algorithms MACPO and MAPPO-Lagrange~\citep{gu2021multi}, which are the safe extensions of HATRPO~\citep{kuba2021trust} and MAPPO~\citep{yu2022surprising} respectively. However, these methods are not guaranteed to work under free-form language constraints and are unable to deal with multiple constraints simultaneously. Other research directions include approaches based on the shielding and barrier functions ~\cite{elsayed2021safe,cai2021safe}, but these methods require pre-training or strong prior knowledge to create barriers that filter actions and cannot generalize to new scenarios, and these barrier functions will change if constraints change.

\textbf{Constraints with Natural Language:} Previous works used natural language to constrain agents to behave safely under a single agent setting. \citet{prakash2020guiding} trained a constraint checker in a supervised fashion to predict whether the natural language constraints are violated and guide RL agents to learn safe policies. During training, a ground-truth cost for each constraint was required to train the constraint checker. However, this approach may not be feasible if the constraint or language structure changes during the application. \citet{yang2021safe} trained a constraint interpreter to predict which entities in the environment may be relevant to the constraint and used the interpreter to predict costs. Their approach did not rely on a ground-truth cost, but the interpreter had to model and predict all entities in the environment. This necessitated a constraint in a similar structure, which could result in inaccurate outcomes in complex tasks since the cost prediction model cannot handle free-form language. Our method, in contrast, utilizes Language Models to predict constraint violations, eliminating the need for ground-truth costs and extra training modules.

\textbf{Language Models: } 
In recent years, LMs based on utilizing Transformers~\citep{vaswani2017attention} have attracted great attention. For example, Bidirectional Encoder Representations from Transformers (BERT)~\citep{devlin2018bert} focuses on extracting semantic meaning and learning representations for text inputs by joint conditioning on their context, which can be easily fine-tuned for downstream tasks. Models such as the GPT~\citep{brown2020language} and Llama~\cite{touvron2023llama} have been developed to generate text by incorporating extensive prior knowledge with an emphasis on the decoder aspect. These models are trained to create text based on the preceding context and have shown proficiency in text-generation tasks. As Language Models provide the potential to align human language with policy learning and decision-making domains, previous research has attempted to introduce Language Models into MARL ~\cite{chen2023agentverse}. However, to the best of our knowledge, our work is the first to apply the fine-tuned LMs to the field of Safe MARL specifically to tackle natural language constraint challenges.

\section{Preliminaries}

\textbf{Constrained Markov Game}~\citep{altman2021constrained,gu2021multi} is defined by a tuple $\langle N, S,A,P,R,\gamma, C, d \rangle $, where $N = \{1, ... , n\}$ is the set of agents, $S$ is the state space, $A$ is the action space, $P: {S} \times {A} \times {S} \rightarrow \mathbb{R}$ is the probabilistic transition function, $R: {S} \times {A} \rightarrow \mathbb{R}$ is the team reward function, $C: {S} \times {A} \rightarrow \mathbb{R}$ is the set of cost functions, $d$ it the constraint violation budget and $\gamma \in[0,1)$ is the discount factor. At time step $t$, the agents are in state $s_t$, and each agent $i$ chooses an action $a^i_t$ according to its policy $\pi^i\left({a}^i \mid \mathbf{s}_t\right)$. The joint action represented by $\mathbf{a}_t=\left({a}_t^1, \ldots, {a}_t^n\right)$, and $\boldsymbol{\pi}(\mathbf{a} \mid \mathbf{s})=\prod_{i=1}^n \pi^i\left(\mathbf{a}^i \mid \mathbf{s}\right)$ denotes joint policies. All agents will receive the team reward $r_t$ and the cost $c^i_t$ for each agent. In this paper, we consider the fully-cooperative setting, where all agents aim to maximize the expected team reward,
\begin{equation}\label{reward_return}
J_r(\boldsymbol{\pi}) \triangleq \mathbb{E}_{s_0 \sim \rho^0, \mathbf{a}_{0: \infty} \sim \boldsymbol{\pi}, s_{1: \infty} \sim {P}}\left[\sum_{t=0}^{\infty} \gamma^t R\left( s_t, \mathbf{a}_t\right)\right]
\end{equation}
and minimize the accumulated cost by simultaneously satisfying the constraints
\begin{equation}\label{cost_return}
J _c(\boldsymbol{\pi}) \triangleq \mathbb{E} _{\mathrm{s} _0 \sim \rho^0, \mathbf{a} _{0: \infty} \sim \boldsymbol{\pi}, \mathrm{s} _{1: \infty} \sim P}\left[\sum _{t=0}^{\infty} \gamma^t C\left(\mathrm{~s} _t, \mathbf{a} _t\right)\right]
\end{equation}
The objective of the \textit{constrained Markov game} is to find the optimal joint policy $\boldsymbol{\pi}^*$ that maximizes the expected team reward while satisfying the cost constraints, i.e.,
\begin{equation}\label{cmdp_obj}
\boldsymbol{\pi}^* = \arg\max_{\boldsymbol{\pi}} J_r(\boldsymbol{\pi}), \text{ s.t. } J_c(\boldsymbol{\pi}) \leq d.
\end{equation}
In the traditional Constrained Markov Game formulation, the cost function $C$ plays a crucial role in quantifying the degree of constraint violation. However, this predefined cost function has limitations in practice, such as requiring extensive domain knowledge for design and lacking flexibility to adapt to dynamic and unstructured constraints.

\section{Methodology}
\label{sec:method}
In this section, we introduce the Language Constrained Markov Game and present our safe MARL method called SMALL, which consists of a Cost Learning Module to anticipate constraint violations using Language Models and a Multi-Agent Policy Network for action generation based on environment observations and insights from cost learning.

\subsection{Language Constrained Markov Game}
We consider the problem where the cost function $C$ is not known but instead derived from some free-form natural language. Thus, instead of a Constrained Markov Game, we model the problem as the tuple $\langle N,S,A,P,R,\gamma, P_c, L ,C, d \rangle$. In contrast to the Constrained Markov Game, this Language Constrained Markov Game is augmented by a constraint transformation function $P_c$ and natural language constraint space $L$, where $P_c: L\rightarrow C^l$ maps some natural language constraint $l\in L$ to a cost function $C^l$, where $C^l:S\times A\rightarrow \{0,1\}$ decides whether the agent has violated the natural language constraints. Under this setting, agents only know the natural language constraint $l$ but lack the knowledge of the ground-truth cost $C^l(s_t,\mathbf{a}_t)$. In this paper, we sample the natural language constraint $l$ at the beginning of each episode and use it throughout the subsequent training phase. We assume that each $l$ corresponds to a cost function $C^l$. However, $l$ may contain redundant or irrelevant information. Therefore, we introduce a simplified version of the constraint, denoted as $l_c$, which is obtained by removing the redundant information from $l$. We will use the variable $l_c$ to refer to these simplified natural language constraints in the rest of the paper.

\subsection{Cost Learning Module}
\label{sec:cost_pred}
%As the framework illustrated in Figure~\ref{fig:struc}, the model processes natural language instructions by first channelling the language input $l$ through a decoder language model. This step is crucial for condensing the verbose and potentially semantically ambiguous free-form language constraints into a more concise representation $l_c$. To efficiently manage this complexity and ensure a clear understanding of the constraints, we utilize the GPT~\cite{brown2020language} model for this induction phase, as it aligns well with human values and is adept at disambiguating and summarizing the essence of natural language constraints. Following this, we utilize BERT~\cite{devlin2018bert}, an encoder-decoder language model, to extract semantic information from the compressed natural language constraint $l_c$ and convert it into a constraint embedding $E_l$. To align compacted language constraints $l_c$ with embeddings $E_l$, we fine-tune the encoder language model using a triplet loss, which is particularly effective in optimizing embedding spaces for superior discriminative properties. The loss for a set of natural language constraint pairs can be defined as follows:
We design a cost learning module to convert natural language descriptions into costs for safety control. The first step involves summarizing natural language instructions by channelling the language input $l$ through a large language model (LLM). This step is crucial for condensing verbose and potentially semantically ambiguous free-form language constraints into a concise representation $l_c$. To efficiently manage this complexity and ensure a clear understanding of the constraints, we leverage the LLM for this induction phase. LLMs, such as GPT-3.5~\cite{brown2020language}, align well with human values and are adept at disambiguating and summarizing the essence of natural language constraints.
Following this, we train a model to extract semantic information from the natural language constraint $l_c$, converting it into a constraint embedding $E_l$. To learn these embeddings $E_l$, we introduce a model based on BERT~\cite{devlin2018bert}, an encoder-decoder language model, and fine-tune it using contrastive learning. We use triplet loss, where a positive and a negative sample are simultaneously taken as input with the anchor sample, defined as follows:
\begin{equation}
\label{eq.finetune}
    \mathcal{L}_{\text {tri }} = \frac{1}{n} \sum_{k=1}^n[\max (0, \alpha+\operatorname{dist}({E_{l_1}^k}, {E_{l_2}^k})-\operatorname{dist}({E_{l_1}^k}, {E_{l_3}^k}))]
\end{equation}
where ${E_{l_1}^k}$, ${E_{l_2}^k}$ and ${E_{l_3}^k}$ represent the embeddings of the $k$-th natural language constraint triplet $({l_1}^k,{l_2}^k,{l_3}^k)$, the $\alpha$ is the margin term ensures a minimum separation between positive and negative examples in the embedding space. Particularly, $E_{l_1}^k$ is an embedding of ${l_1}^k$, the positive sample ${E_{l_2}^k}$ is an embedding of ${l_2}^k$ that prohibits the agents from the same entities or behaviour to the ${l_1}^k$, and the negative sample ${E_{l_3}^k}$ is an embedding of ${l_3}^k$ that is different from or unrelated to ${l_1}^k$. The \(\text{dist}(\cdot, \cdot)\), which measures the distance between embeddings, is calculated using cosine similarity in our method. This encourages the model to learn embeddings where similar constraints are closer together while dissimilar ones are farther apart, aligning with the compacted language constraints more effectively. All the mentioned embeddings are generated by encoder and decoder language models, based on natural language constraints. The triplet loss helps the encoder language model to recognize the semantic similarity of constraints ~\cite{reimers2019sentence}. Constraints about the same entities and behaviours will have embeddings with high cosine similarity and vice versa.

After receiving text-based observations for each agent, the encoder language model encodes them into observation embeddings $\bm{E}_{o,t} = \{E^1_{o,t},..., E^n_{o,t}\}$, which capture the semantic essence of the circumstances surrounding each agent $i$ at each given timestep $t$. The embedding $E_l$ is integrated with the environment description to detect any constraint violations. Since $E_l$ represents a concise semantic embedding, it necessitates refining each agent's raw text observation to align more precisely and accurately with the environment description. This refinement enables the prediction of costs using the constraint's semantic embedding. Our method employs a descriptor that automatically filters the general representation related to entities or obstacles, ensuring that only the most pertinent information is considered. This automated filtration significantly optimizes the process of detecting and addressing constraint violations.

To determine the cost of violating constraints, we first calculate the cosine similarity between the constraint embedding $E_l$ and the observation embeddings $E_{o,t}^i$, denoted as $\text{sim}(E_l, \bm{E}_{o,t}) \in [0, 1]^{n}$. However, we find that relying solely on this similarity score may lead to an insufficient understanding of the constraints. To address this issue, we introduce an additional step that queries a decoder language model with the current text-based observation and the language constraint as a prompt, asking whether the constraint has been violated. The decoder outputs a binary flag $v^i_t \in {0, 1}$, where $v^i_t = 1$ indicates that agent $i$ has violated the constraint at timestep $t$, and $v^i_t = 0$ otherwise. We then multiply the cosine similarity $\operatorname{dist}(E_l, E_{o,t}^i)$ with $v^i_t$ to obtain the final predicted cost $\hat{c}^i_t$ for agent $i$:
\begin{equation}\label{eq:c_predict}
\hat{c}^i_t = v^i_t \cdot \operatorname{dist}(E_l, E_{o,t}^i), \text{ for } i \in N.
\end{equation}
For the validation query, we utilize LLMs, such as Llama3-8B~\citep{touvron2023llama}. This approach leverages the decoders' capability to determine constraint violations, despite their potential difficulty in explicitly outputting cost values. By combining the strengths of cosine similarity and the decoders' binary output, we achieve a more accurate and informative cost prediction mechanism, capitalizing on the complementary abilities of these two components.

\subsection{Multi-Agent Policy Learning with Constraints}

After obtaining the predicted cost $\hat{C} = \{\hat{c}^1_t, ...,\hat{c}^n_t \}$ from the cost learning module, we are ready to train the policy $\boldsymbol{\pi}$ for safe MARL agents. It is worth noting that our method does not require the ground-truth cost under any circumstances for training or evaluation. This feature distinguishes it from other safe MARL algorithms. We integrate the cost learning module to the Multi-Agent Proximal Policy Optimization (MAPPO)~\cite{yu2022surprising} and Heterogeneous-Agent Proximal Policy Optimisation (HAPPO) ~\cite{kuba2021trust} with the Lagrange multiplier \cite{ray2019benchmarking}. This allows the agents to maximize their rewards while adhering to specific constraints at the same time.

% For a more detailed definition of those functions, please refer to Appendix \ref{app:marl}.

Drawing an analogy to the return function $J_r(\boldsymbol{\pi})$ in Equation \ref{reward_return}, the value function $V_{\boldsymbol{\pi}}(s)$ and the advantage function $A_{\boldsymbol{\pi}}^i\left(s, \boldsymbol{a}^{i}\right)$, we can define corresponding cost-related functions. Specifically, we introduce the cost return function $J_c$, state cost value function $V_{c, \boldsymbol{\pi}}^i(s)$ and cost advantage function $A_{c}^i(s, a^i)$. The joint policy $\boldsymbol{\pi}$ can be obtained by
\begin{equation}\label{equ:pi}
\boldsymbol{\pi}=\argmax\limits_{\boldsymbol{\pi}} J_r(\boldsymbol{\pi})-\lambda J_{c}(\boldsymbol{\pi}),
\end{equation}
where $J_{c}(\boldsymbol{\pi}) =  \mathbb{E}_{\boldsymbol{\pi}} \left[\sum_{t=0}^{\infty} \gamma^t \sum^N_i\hat{c}^i_t\right]$ is the expected cost sum of all agents, where $\hat{c}^i_t$ is the predicted cost for agent $i$ at timestep $t$, and $\lambda$ is the Lagrange multiplier.
The training of the value function $V_{\boldsymbol{\pi}}(s)$ and cost value function $V_{c, \boldsymbol{\pi}}^i(s)$ is updated by minimizing the corresponding mean squared TD-error as
\begin{align}
\label{value_function}
    \mathcal{L}^{v}&=\mathbb{E}_{\boldsymbol{\pi}}\left[\left(R_t+\gamma V_{\boldsymbol{\pi}}(s_{t+1})-V_{\boldsymbol{\pi}}(s_t)\right)^2\right],\\
    \mathcal{L}^{v}_c&=\mathbb{E}_{\boldsymbol{\pi}}\left[\frac{1}{2}\left(\hat{c}_t^i+\gamma V_{c,\pi}^i(s_{t+1},E_l)-V_{c,\pi}^i(s_t,E_l)\right)^2\right],
\label{value_function_c}
\end{align}
where $E_l$ is the constraint embedding from the encoder language model, $\hat{c}^i_t$ is the predicted cost for agent $i$. To maximize the return $J_r$ and minimize the cost $J_c$, we can adapt the PPO-clip objective ~\cite{schulman2017proximal} to update the policy with first-order methods. Building on this framework, we seamlessly integrate this approach with the MARL algorithm, specifically leveraging the PPO-based objective updates to facilitate policy learning. As a result, we utilize HAPPO~\cite{kuba2021trust} and MAPPO~\cite{yu2022surprising} as the backbones to develop the \textbf{SMALL-HAPPO} and \textbf{SMALL-MAPPO} algorithms, respectively. These algorithms are then benchmarked against other baselines in the subsequent experimental section, with the proposed method's pseudo-code detailed in Algorithm \ref{shortalgo}.

\begin{algorithm}[H]
    \caption{ \textbf{S}afe \textbf{M}ulti-\textbf{A}gent \textbf{R}einforcement \textbf{L}earning with natural \textbf{L}anguage constraints (\textbf{SMALL})}
\begin{algorithmic}[1]
    \STATE Fine-tune the language constraint encoder $\mathbf{E}$ (Eq. \ref{eq.finetune}).
    \FOR{each episode}
    \STATE Sample language constraints $l \in L$.
    \STATE Use the LLM to condense $l \rightarrow l_c$; Use the language constraint encoder $\mathbf{E}$ to encode $l_c \rightarrow E_l$.
    \FOR{agent $i = 1, \dots, n$}
    \STATE Roll-out the policy with $E_l$ and get trajectory $\{o^i_t,a^i_t,{o^i_t}^\prime,r_t\}_{t=1,..,T}$.
    \FOR{$t = 1, \dots, T$}
    \STATE Transform $o^i_t \rightarrow \hat{o}^i_t$ into compact environment description.
    \STATE Encode $\hat{o}^i_t \rightarrow E_{o,t}^i$ using the language constraint encoder $\mathbf{E}$
    \STATE Get the violation $v^i_t$ through the LLM.
    \STATE Predict cost $\hat{c}^i_t$ (Eq. \ref{eq:c_predict}). 
    \ENDFOR
     \STATE Update policy $\boldsymbol{\pi}$ 
 (Eq.~\ref{equ:pi},\ref{value_function},\ref{value_function_c} and HAPPO / MAPPO)
    \ENDFOR
    \ENDFOR
\end{algorithmic}
\label{shortalgo}
\end{algorithm}

\section{LaMaSafe Benchmark}
\label{sec:lamasafe}

\begin{figure*}[t]
\centering  %居中
\subfigure[LaMaSafe-Grid: Random Layout]{   %第一张子图
\begin{minipage}{0.23\textwidth}
\centering    %子图居中
\includegraphics[width=\textwidth]{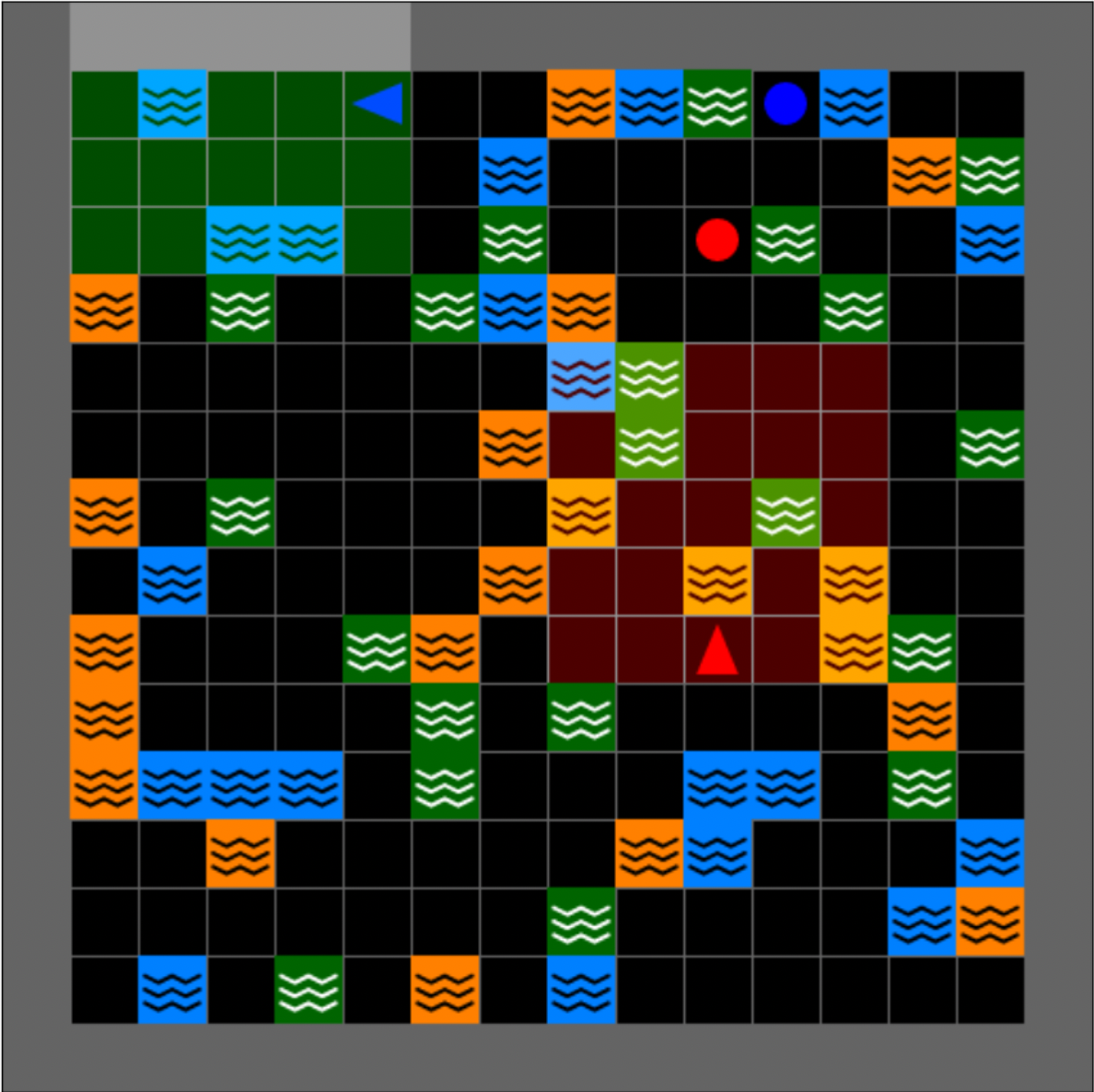}   %以pic.jpg的0.5倍大小输出
\end{minipage}
}
\subfigure[LaMaSafe-Goal (Ant): Hard Layout, 32H/10V]{ %第二张子图
\begin{minipage}{0.23\textwidth}
\centering    %子图居中
\includegraphics[width=\textwidth]{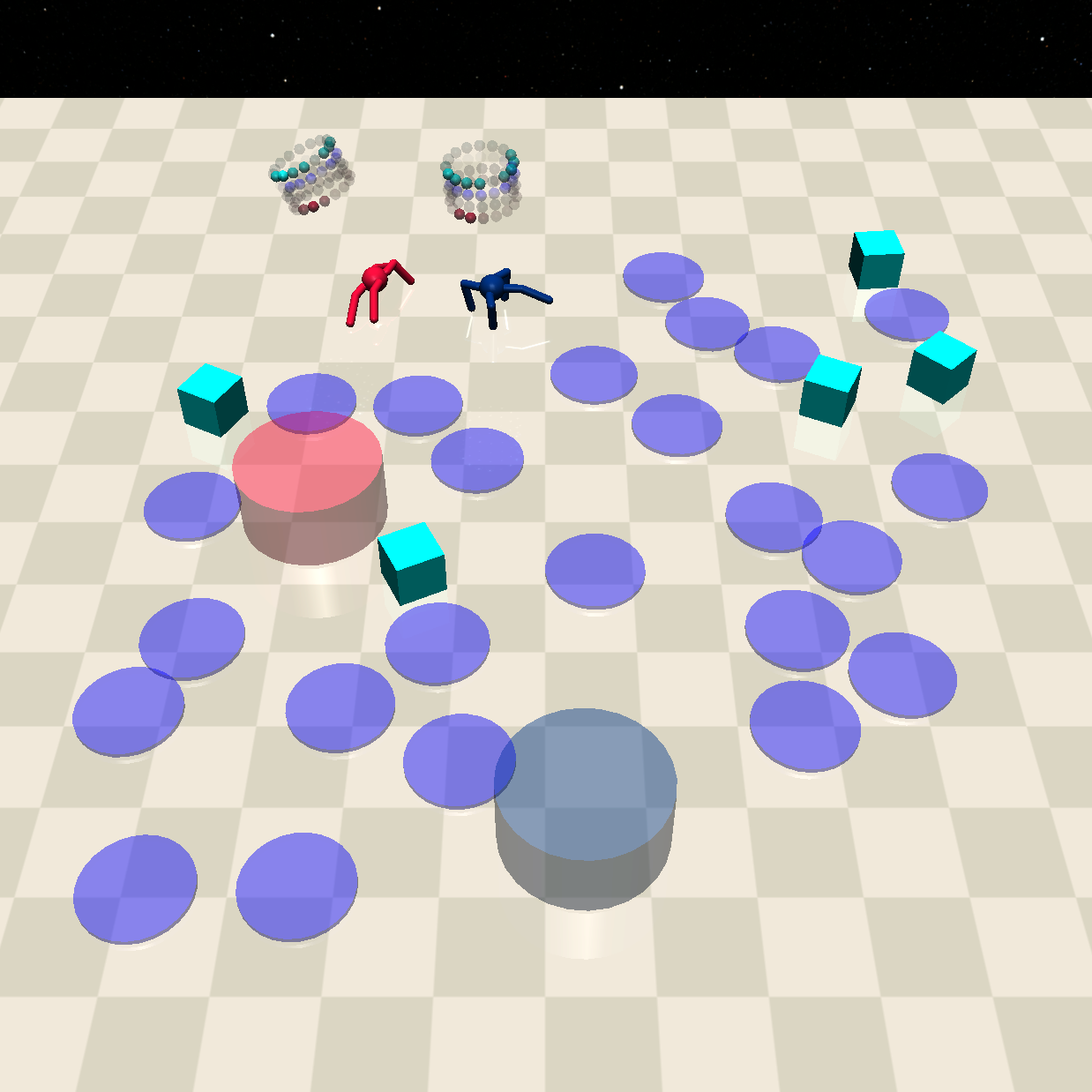}%以pic.jpg的0.5倍大小输出
\end{minipage}
}
% 第三张和第四张子图，上下排列
\begin{minipage}{0.35\textwidth}
\vspace{0.7cm}
  \subfigure[Natural Language Constraints]{ 
\centering    %子图居中
\resizebox{\textwidth}{!}{%
\begin{tabular}{|c|}
\hline
Robots must steer clear of any blue circles in the area. \\ \hline
You can't be too close to each other. \\ \hline
The agents must not dance in the water lest they rust. \\ \hline
The agents should not follow the path of the lava \\ \hline
Blue circles are to be considered danger zones – avoid them \\ \hline
... \\ \hline
\end{tabular}
}
  }

\subfigure[Environment Description]{ 
  \centering    %子图居中
\resizebox{\textwidth}{!}{%
\begin{tabular}{|c|}
\hline
Hazards have been detected within 3m, 2.5m and 1m of you.\\ \hline
No hazards were detected. There are \textbf{one vase} close to you.\\ \hline
There are \textbf{three hazards}close to you, in 1.5m, 4m and 4.5m.\\ \hline
There is lava on your right and water on your back. \\ \hline
There is grass on your left and water on your back. \\ \hline
There is noting around you. \\ \hline
... \\ \hline
\end{tabular}
\label{table:environment_description}
}
  }
\end{minipage}
\caption{\textbf{LaMaSafe Benchmark.} (a) Grid: two agents in \texttt{Random} layout, including 20 randomly placed lava, water and grass. (b)Goal(Ant): two agents in the \texttt{Hard} level layout, in which each agent controls the four joints of an ant to navigate. The numbers behind indicate the obstacle count, in which ``H'' and ``V'' represent hazards and vases, respectively. The task's difficulty level increases with the number of hazards and vases. (c) Examples of natural language constraints employed in our evaluation. (d) Examples of environmental descriptions provided by the environments.}
\label{fig:envs_main}  
\end{figure*}

Although researchers in the field of safe MARL have access to a diverse range of environments for testing various algorithms such as Safe MAIG~\cite{gu2023safe}, Safety-Gymnasium~\cite{ji2023safety} and SMAMuJoCo~\cite{gu2021multi}, there remains a notable gap in the availability of safe MARL environments that incorporate natural language constraints. As a side contribution of this paper, we propose a new language-based constraint safety multi-agent environment named \textit{LaMaSafe}, which contains two types of environments, namely the LaMaSafe-Grid and  LaMaSafe-Goal. As shown in Figure~\ref{fig:envs_main}, LaMaSafe-Grid is a 2D discrete-action environment based on the Mini-Grid~\citep{MinigridMiniworld23} with human language safety constraints, while LaMaSafe-Goal is a 3D continuous-action environment based on the Gymnasium~\citep{towers_gymnasium_2023}. In both environments, agents are required to navigate and complete objectives while adhering to constraints expressed in free-form natural language, such as "do not pass through the lava area" or "avoid collisions between two agents".

\textbf{LaMaSafe-Grid. }is a 2D multi-agent navigation environment where agents operate in a grid world, aiming to find their designated target balls. Each agent has its own ball, and finding it yields a reward of 3 points that linearly decays to 0.1 times its original value over the course of an episode. The environment features three types of hazardous areas: lava, water, and grass. At the beginning of each episode, a natural language constraint informs the agents about the specific hazardous area they must avoid. Agents must also maintain a safe distance from each other to prevent collisions, which incur penalties and increase the ground-truth cost. The objective is to maximize rewards while minimizing constraint violations, which are tallied at the end of each episode. An episode terminates when all agents have found their respective balls or when the maximum number of timesteps (set to 300) is reached. The environment provides two layouts: (1) Random, where hazards are randomly scattered throughout the grid world, and (2) One-Path, where the entire grid world is filled with lava, and each agent has only a single safe path to navigate, featuring numerous turns. The difficulty of the One-Path layout lies in the fact that if the constraint requires avoiding lava, the agents can only traverse the safe path, while in other cases, they have the freedom to move freely. LaMaSafe-Grid incorporates both hazardous area avoidance and inter-agent collision prevention, requiring careful coordination and planning among the agents to optimize their performance. Appendix~\ref{app:lamasafe-grid} provides additional details of the environment setting.

\textbf{LaMaSafe-Goal. }is fundamentally a linguistic 3D multi-agent environment where agents control three types of robots navigating upon a plane, namely the Point, Car and Ant. As a safety environment, it features three kinds of constraints: 1) Hazards, which are non-contact flat blue circles; 2) Vases, which are contactable pale green cubes; and 3) Collisions, where the two robots get too close to crash. There are three difficulty levels, each corresponding to different quantities of hazards and vases.
As a highlight, the environment is enriched with human-described natural language constraints, such as \textit{``Robots must steer clear of any blue circles in the area,''} or \textit{``Agents will be injured when they collide with each other and avoid crashing ''}. The robots aim to reach their designated target locations while adhering to the natural language constraints specified at the start of each episode. Once a robot arrives at its target location, its goal is randomly relocated to an unoccupied position. The ground truth cost utilized in evaluating the experiments' performance is dictated by the frequency of natural language constraint violations. To simplify the cost calculation process, we designed all the natural language constraints to prioritize avoiding hazards in both simple and abstract settings. Therefore, the cost metric in the experiments is the number of times agents come into contact with hazards. Episodes terminate upon reaching the maximum time step, which is set to 1000 in our experiments. Please refer to Appendix~\ref{app:lamasafe-goal} for detailed descriptions and further elaboration on layouts.

% \begin{wrapfigure}[16]{r}{0.5\textwidth}
% \vspace{-0.6cm}
%     \begin{center}\includegraphics[width=\linewidth]{img/4agent_ablation.pdf}    \vspace{-0.5cm}
%         \caption{\textbf{Four Agent Comparison}: SMALL with MAPPO and
% HAPPO on the Easy, Hard level of LaMaSafe-Goal(Ant) involving four agents.}
%         \label{fig:4agents}
%     \end{center}
% \end{wrapfigure}

\section{Experiments}
\label{sec:exp}
In this section, we evaluate our method using the LaMaSafe-Grid and LaMaSafe-Goal benchmarks. We conduct experiments in various multi-agent environments. Our objective is to validate that our method, SMALL, works efficiently for safe MARL with language constraints.

\begin{figure*}[t] \centering\includegraphics[width=\textwidth]{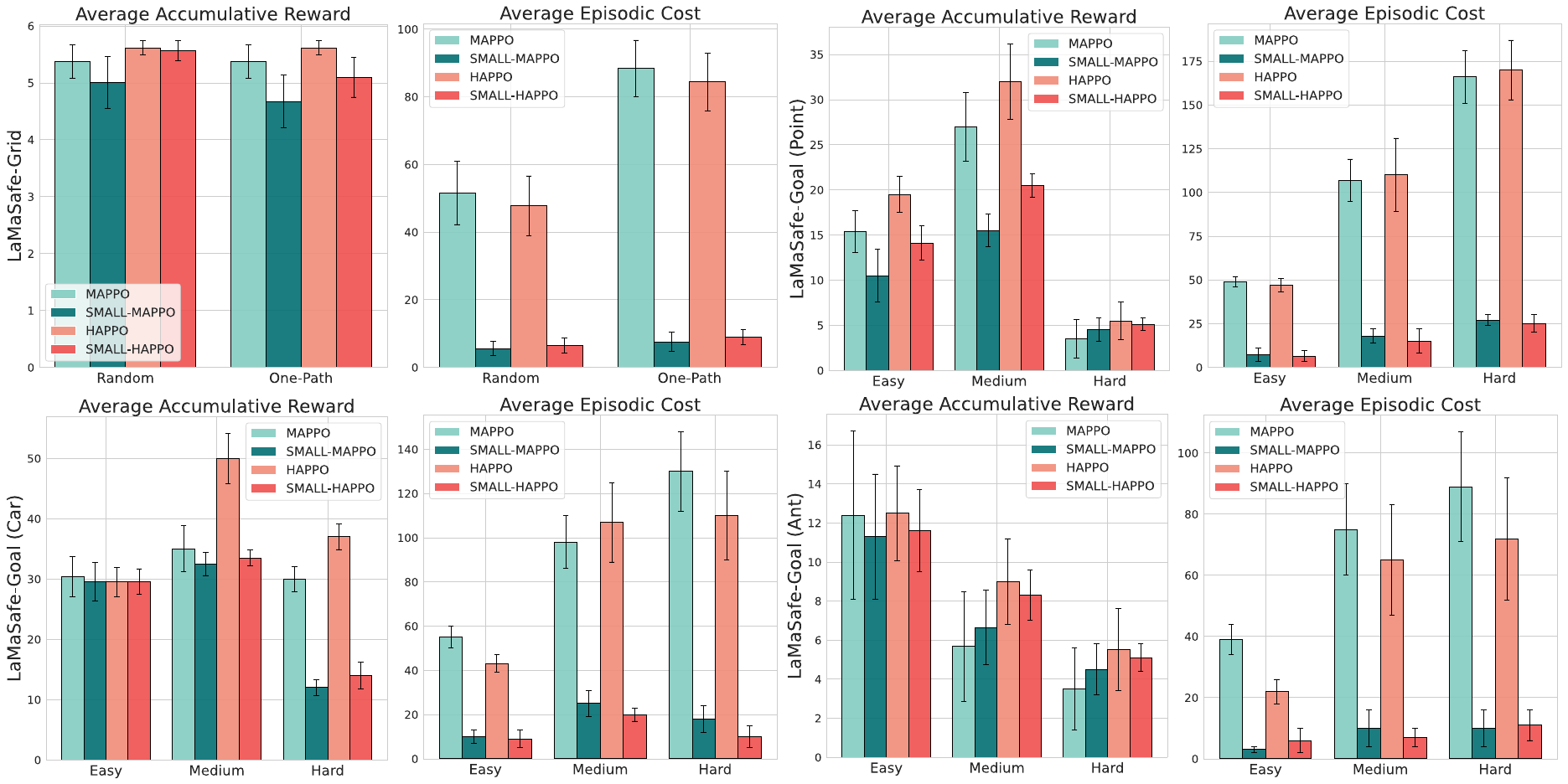}
\vspace{-0.45cm}
    \caption{\textbf{Comparison in Natural Language Constraints:} We conducted a comparison of the performance of four different algorithms, namely MAPPO, HAPPO, SMALL-MAPPO, and SMALL-HAPPO in LaMASafe-Grid and LaMASafe-Goal. The evaluation was based on rewards and costs across different types of agents and layouts. It is important to note that the comparison of all algorithms only takes into account natural language constraints. To ensure a fair comparison, we augmented the embedding $E_l$ to the state for MAPPO and HAPPO.}
    \label{fig:main}
\vspace{-0.45cm}
\end{figure*}

\subsection{Setup}
%\textbf{Baselines.} We compare our method with four baselines,  \textbf{MAPPO}~\cite{yu2022surprising} including an algorithm that scales PPO to multi-agent systems by employing centralized training with decentralized execution \textbf{HAPPO} ~\cite{kuba2021trust} which introduces a trust region method tailored for heterogeneous agent policies, \textbf{MAPPO-Lagrange}~\cite{gu2021multi} an extension of the MAPPO framework that integrates a Lagrangian approach to dynamically adjust constraints, thereby ensuring safer policy updates in environments with the pre-defined cost function and \textbf{Happo-Lagrange} designed as an extension of HAPPO by mimicking the MAPPO-Lagrange approach. For more details, please refer to Appendix \ref{app:imp}.
\textbf{Baselines.} We compare our method with four baselines: MAPPO~\cite{yu2022surprising}, an algorithm that scales PPO to multi-agent systems by employing centralized training with decentralized execution; HAPPO~\cite{kuba2021trust}, which introduces a trust region method tailored for heterogeneous agent policies; MAPPO-Lagrange~\cite{gu2021multi}, an extension of the MAPPO framework that integrates a Lagrangian approach to dynamically adjust constraints, thereby ensuring safer policy updates in environments with a pre-defined cost function; and HAPPO-Lagrange, designed as an extension to HAPPO by mimicking the MAPPO-Lagrange approach. For more details, please refer to Appendix~\ref{app:imp}.

\textbf{Metrics.} We assess the algorithms' capability to adhere to human-provided natural language constraints while maximizing rewards. This evaluation is conducted by measuring the average reward obtained across three random seeds, under the condition that the agents follow the specified natural language constraints. Intuitively, the ability to secure higher rewards under these constraints signifies more effective compliance and understanding of the natural language directives.

\subsection{Main Results} 
We demonstrate the performance of algorithms in an environment solely guided by natural language constraints, comparing MAPPO and HAPPO with our methods, SMALL-MAPPO and SMALL-HAPPO. As shown in Figure \ref{fig:main}, we present the learning curves for rewards and costs across different LaMaSafe environments and tasks, with our proposed algorithms represented by the deep green and red boxs. From a reward perspective, due to the need to consider various natural language constraints, SMALL-based algorithms generally perform slightly below their backbone algorithms. However, they maintain a similar level of performance and excel in more challenging scenarios, such as the Ant Medium and Hard layeouts. In terms of cost, a significant difference is observed. MAPPO and HAPPO struggle to handle natural language constraints and often incur high costs. In contrast to other algorithms, SMALL-based algorithms are highly efficient and converge to extremely low cost in almost all environments. This proves that SMALL is not only capable of understanding natural language descriptions that it has never encountered before but also ensures that constraints are adhered to, leading to maximum team rewards while minimizing the number of violations.

\subsection{Ablation Study}

\begin{figure*}[t] \centering\includegraphics[width=\textwidth]{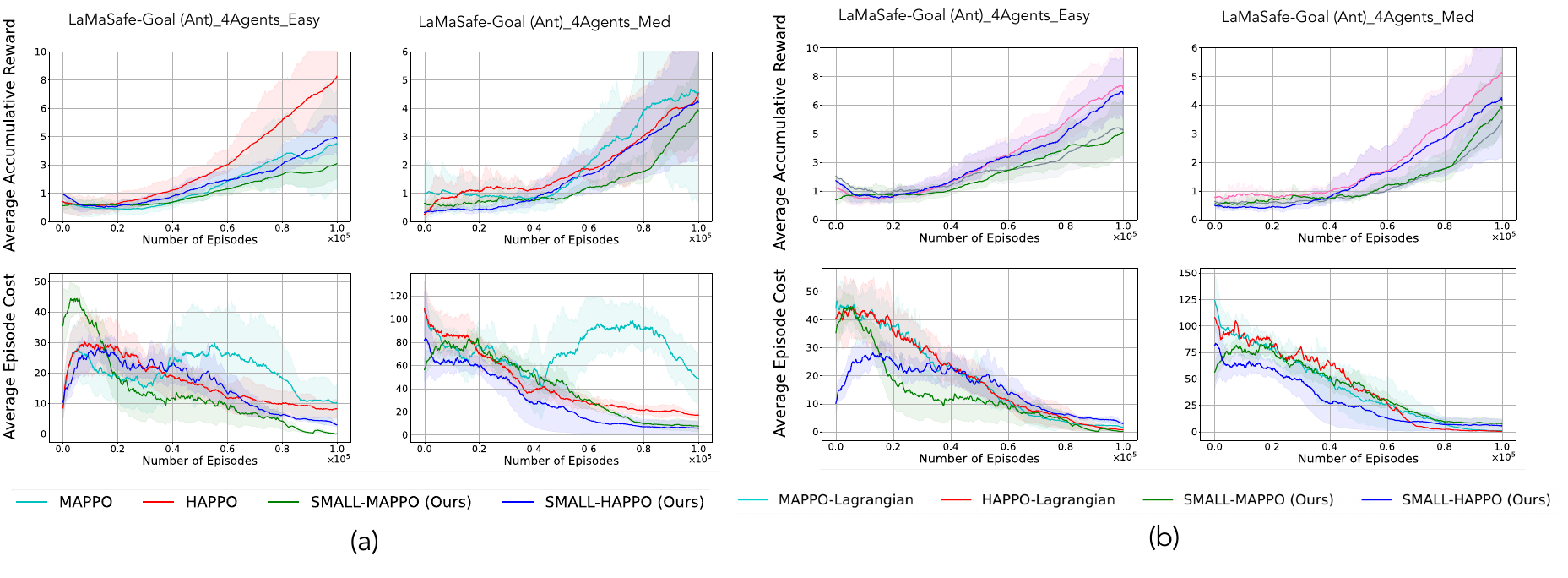}
\vspace{-0.73cm}
    \caption{(a) \textbf{Four Agent Comparison}: SMALL with MAPPO and HAPPO on the Easy, Hard level of LaMaSafe-Goal(Ant) involving four agents. (b) \textbf{Ground Truth Cost Comparison}: SMALL with MAPPO-Lagrange and
HAPPO-Lagrange on the Hard level of Goal(Ant) with four agents.}
    \label{fig:4agents}
 \vspace{-0.45cm}
\end{figure*}

\textbf{Scalability to More Agents.} To further investigate SMALL's performance with an increased number of agents, we extended the LaMaSafe-Goal (Ant) environment from the main experiment to include 4 agents, doubling the number of agents compared to the original setup. The natural language constraints remained focused on avoiding blue hazards and preventing collisions, which significantly increased the complexity and difficulty of the environment. Figure~\ref{fig:4agents} (a) presents the results of this extended experiment. Consistent with the main results, the SMALL algorithm maintains rewards slightly lower than the baseline while substantially reducing the number of constraint violations. However, in 4-agent Easy and Medium layouts, the baseline algorithms (particularly HAPPO) exhibit a convergence trend in cost performance. We hypothesize that augmenting the constraint embedding $E_l$ within the state representation may contribute to this behavior, drawing inspiration from the findings of \cite{sootla2022saute}, which demonstrate that augmenting the state input of the policy network facilitates the understanding of natural language constraints.

\textbf{Comparison with algorithms that use the ground truth cost.} We compare our algorithms with Safe MARL methods that use the ground truth costs for learning. This comparison explores our method's precision in learning from natural language descriptions. As shown in Figure~\ref{fig:4agents} (b), in terms of cost, our algorithms converge similarly, demonstrating consistent performance even in the most challenging environments. Algorithms that have access to ground truth costs are considered oracles in this context. Consequently, most SMALL-based algorithms perform similarly or slightly worse.

It is noteworthy that our algorithms occasionally outperform others. We conjecture that this may stem from the cost prediction module incorrectly considering certain high-risk yet potentially beneficial actions (such as navigating close to hazardous zones) as acceptable. This suggests that our approach excels at understanding natural language instructions and may encourage bold strategies by identifying opportunities for reward, even if it means breaking the rules.

% General results on all five tasks are given in Fig. \ref{fig:main_res_!} and \ref{fig:main_res_2}. We report episode reward sum and episode cost sum, respectively to show how the methods perform. As we use cost prediction (CP) to replace ground-truth costs, we refer to our method as PPO\_CP. 

\begin{wrapfigure}[10]{r}{0.6\textwidth}
\centering
\vspace{-0.2cm}
\begin{minipage}{0.6\textwidth}
\begin{table}[H]
\caption{Ablations on SMALL components in LaMaSafe-Goal(Ant) with the 2 agent Easy Layout.}
\begin{tabular}{ccc}
\hline
Ablations                                  & Reward       & Cost         \\ \hline
\multicolumn{1}{c|}{w/o Fine-tuning (Eq.\ref{eq.finetune})} & 6.33±3.42    & 10.45±3.94   \\ \hline
% \multicolumn{1}{c|}{Using LSTM as ${LM_e}$}  & 1.45±0.67    & 20.39±7.26    \\ \hline
\multicolumn{1}{c|}{w/o Decoder}  & 12.50±2.50    & 10.75±3.65    \\ \hline

\multicolumn{1}{l|}{w/o Descriptor (App.\ref{app:des} )}    & 7.89±3.10    & 9.62±4.01    \\ \hline
%\multicolumn{1}{c|}{Abstract $L$ (App.\ref{app:abs_nlc} )}          & 4.72±4.12   & 17.16±6.20   \\ \hline
\multicolumn{1}{c|}{w/o $v^i_t$ (Eq. \ref{eq:c_predict})}  & 5.12 ± 1.46     & 4.78 ± 1.07    \\ \hline
\multicolumn{1}{c|}{\textbf{SMALL-HAPPO}}  & 11.62±2.13   & 5.82±4.24    \\ \hline
\end{tabular}%
% }
\vspace{0.5cm}
\label{tab:abl-table}
\end{table}
\end{minipage}
\end{wrapfigure}

\paragraph{Ablation on SMALL components.} To assess the effectiveness of each component within SMALL, we performed an ablation study using SMALL-HAPPO as the base model, detailed in Table~\ref{tab:abl-table}. The first component analyzed was the fine-tuning process. In the experimental setup, we fine-tune the encoder language model ($LM_e$) by sampling 30 triplets $\left(l_1{ }^k, l_2{ }^k, l_3{ }^k\right)$ from an alternative set $L_{\text{fine-tune}}$, which is distinct from the $L$ set used in subsequent training. This step, conducted over 95 rounds, is critical for aligning the \texttt{Bert} with the semantics of the potential natural language constraints, as outlined in Equation~\ref{eq.finetune}. The absence of this fine-tuning phase leads to inaccurate predictions by the decoder language model, resulting in suboptimal performance. The second ablation examines the impact of removing the decoder language model. Without the decoder, the system depends solely on the encoded representation, leading to performance akin to non-safe algorithms and a marked decrease in constraint adherence efficiency. The third ablation, removing the descriptor, shows that directly encoding redundant textual observation will degrade performance, emphasizing the importance of precise information management for effective constraint adherence. The last ablation, removing the $v^i_t$ from equation~\ref{eq:c_predict}, directly predicts the cost using the similarity score. This ablation leads to more conservative performance, resulting in lower costs and significantly lower rewards.

\section{Conclusion}
\label{sec:conclusion}
In this paper, we introduced SMALL, a novel approach for Safe Multi-Agent Reinforcement Learning with Natural Language constraints. SMALL addresses the challenge of incorporating diverse and context-specific natural language constraints into MARL by utilizing fine-tuned language models to interpret and adhere to these constraints during policy learning. We developed the LaMaSafe benchmark, which provides a pioneering suite of multi-agent environments that incorporate free-form textual constraints, enabling the evaluation of various algorithms under realistic safety scenarios. Empirical results demonstrate that SMALL achieves comparable rewards to other MARL algorithms while significantly reducing constraint violations, highlighting its effectiveness in understanding and enforcing natural language constraints.

While SMALL represents a significant step forward in safe MARL with natural language constraints, there are still limitations to be addressed in future work. One potential direction is to explore the scalability of SMALL to larger multi-agent systems with more agents and complex constraints. Additionally, investigating techniques to handle ambiguous or conflicting constraints could further enhance the robustness of the approach. Despite these limitations, SMALL provides a solid foundation for future research in this exciting and important area of safe multi-agent reinforcement learning.

\bibliography{main.bib}
% for arxiv
\bibliographystyle{abbrvnat}

\newpage
\appendix

\section{Broader Impact Statement}
\label{app:broader_impact}
The research presented in this paper has the potential to make a significant positive impact on the field of safe multi-agent reinforcement learning (MARL) and its real-world applications. By introducing a novel approach, SMALL, which enables MARL agents to understand and adhere to natural language constraints, we contribute to the development of more accessible, adaptable, and safer multi-agent systems. This breakthrough could lead to the wider adoption of MARL in various domains, such as robotics, autonomous vehicles, and industrial automation, where safety and compliance with human-defined constraints are of utmost importance. Our work also paves the way for more intuitive human-agent interaction, as users can specify safety requirements and operational boundaries using natural language, making the technology more user-friendly and controllable.

\section{Reproducibility Statement}
\label{app:reproduct}
To promote transparent and accountable research practices, we have prioritized the reproducibility of our method. All experiments conducted in this study adhere to controlled conditions and environments, with detailed descriptions of the experimental settings available in Section~\ref{sec:exp} and Appendix~\ref{app:lamasafe-goal},\ref{app:lamasafe-grid}. The implementation specifics for all the baseline methods and our proposed SMALL are thoroughly outlined in Section~\ref{sec:method} and Appendix~\ref{app:imp}.

\section{Implementation of the LaMaSafe-Grid}
\label{app:lamasafe-grid}

In this section, we will provide detailed information about LaMasafe-Grid. Firstly, we will introduce the layout we used, along with their state and action space. Secondly, we will discuss the types of obstacles and the pre-defined ground truth cost function. Thirdly, we will show the rule-based implementation of the texted observation.

\begin{figure*}[ht]
\centering  %居中
\subfigure[Random]{   %第一张子图
\begin{minipage}{0.3\textwidth}
\centering    %子图居中
\includegraphics[width=0.95\textwidth]{img/LaMaSafe-Grid.png}   %以pic.jpg的0.5倍大小输出
\end{minipage}
}
\subfigure[One-Path]{   %第一张子图
\begin{minipage}{0.3\textwidth}
\centering    %子图居中
\includegraphics[width=0.95\textwidth]{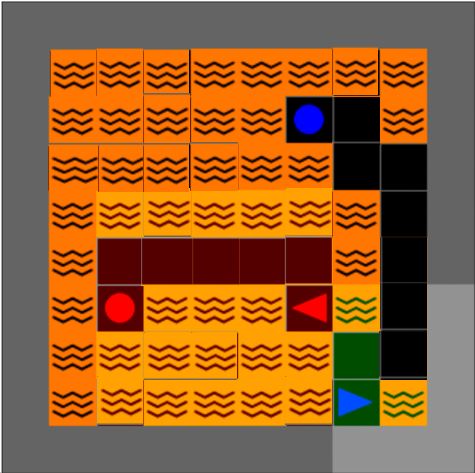}   %以pic.jpg的0.5倍大小输出
\end{minipage}
}

\caption{LaMasafe-Grid, (a) Two agents in Random layout, size 10 by 10, including 20 randomly
placed lava, water and grass. (b) Two agents in One-Path layout,  }   
\label{fig:lamasafe-grid-fig}  
\end{figure*}

\subsection{Layouts}

According to Figure~\ref{fig:lamasafe-grid-fig}, the LaMaSafe-Grid offers two distinct layouts for the agents to navigate: Random and One-Path. 

In the Random layout, the grid-world is a 14x14 matrix, where hazards (lava, water, and grass) are randomly scattered throughout the environment. This layout challenges the agents to adapt to different hazard configurations in each episode while searching for the reward balls.

The One-Path layout, on the other hand, is an 8x8 grid-world filled entirely with lava, except for a single safe path that the agents can traverse. This path is free of hazards but features numerous turns, adding complexity to the agents' navigation. If the natural language constraint requires the agents to avoid lava, they must strictly adhere to the safe path. However, if the constraint allows for more flexibility, the agents can navigate freely within the grid-world.

In both layouts, the agents' objective is to collect balls, each worth a reward of 3 points. The agents must cooperate to maximize their collective reward while adhering to the given natural language constraints. The episode terminates when all balls have been collected or when the maximum number of timesteps (set to 300) is reached.

These contrasting layouts in LaMaSafe-Grid provide diverse challenges for the agents, testing their ability to interpret and follow natural language constraints in different hazard configurations and requiring them to adapt their strategies accordingly.

\subsection{Ground Truth Cost Functions}

\textbf{Collision Detection:} In the grid-world setting, a collision occurs when two agents occupy the same grid cell simultaneously. In such cases, a cost of 1 is assigned to the agents involved in the collision.

\textbf{Hazard Detection:} The cost associated with hazards depends on the human-provided constraints. For example, if the human instructs the agents to avoid lava, then a cost of 1 is assigned whenever an agent occupies a grid cell containing lava.

In both environments, the costs are accumulated over time. If collisions or constraint violations occur across multiple timesteps, the agents will incur cumulative costs proportional to the duration of the violation. This cumulative cost calculation encourages agents to minimize the time spent in violation of the specified constraints.

\subsection{Texted Observation}

In the LaMaSafe-Grid environment, we follow a similar approach to obtain observations and process them before encoding them as embedding vectors.

\textbf{Raw Texted Observation.} The Raw Texted Observation in LaMaSafe-Grid is a simplified version of the one used in LaMaSafetyGoal. Since the grid-world is a discrete environment, the textual descriptions of the state, observation, and action are more concise. The observations include information about the agent's current position, the locations of hazards (lava, water, grass), and the positions of reward balls.

\textbf{Environment Description.} The Environment Description in LaMaSafe-Grid is a rule-based method that processes the Raw Texted Observation to extract relevant information for the agents. Due to the discrete nature of the grid-world, there are only a few possible scenarios:
\begin{itemize}
    \item  \textbf{Agent on a safe tile:} "Agent is on a safe tile. No hazards detected."
    \item  \textbf{Agent on a hazard tile:} "Agent is on a [hazard type] tile. Hazard detected!"
    \item  \textbf{Agent adjacent to a hazard tile:} "Agent is adjacent to a [hazard type] tile. Hazard nearby!"
    \item  \textbf{Agent adjacent to a reward ball:} "Agent is adjacent to a reward ball. Collect the ball!"
\end{itemize}

These concise descriptions provide the agents with essential information about their immediate surroundings, enabling them to make informed decisions based on the presence of hazards and the proximity of reward balls. By leveraging these textual observations, agents can effectively navigate the grid-world while adhering to the given natural language constraints.

The simplified nature of the LaMaSafe-Grid environment allows for a more straightforward application of language in describing the agents' observations and spatial relationships, demonstrating the versatility of the texted observation approach in both continuous and discrete multi-agent settings.

\section{Implementation of the LaMaSafe-Goal}
\label{app:lamasafe-goal}

\begin{figure*}[ht]
\centering  %居中
\subfigure[Point Agent]{   %第一张子图
\begin{minipage}{0.3\textwidth}
\centering    %子图居中
\includegraphics[width=0.95\textwidth]{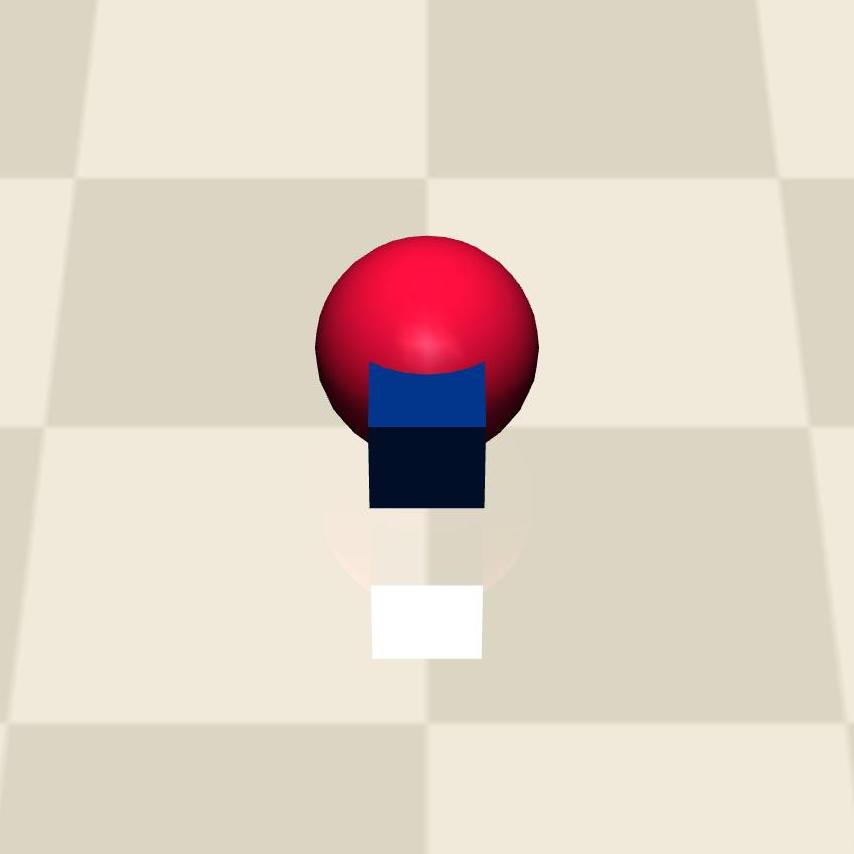}   %以pic.jpg的0.5倍大小输出
\end{minipage}
}
\subfigure[Car Agent]{   %第一张子图
\begin{minipage}{0.3\textwidth}
\centering    %子图居中
\includegraphics[width=0.95\textwidth]{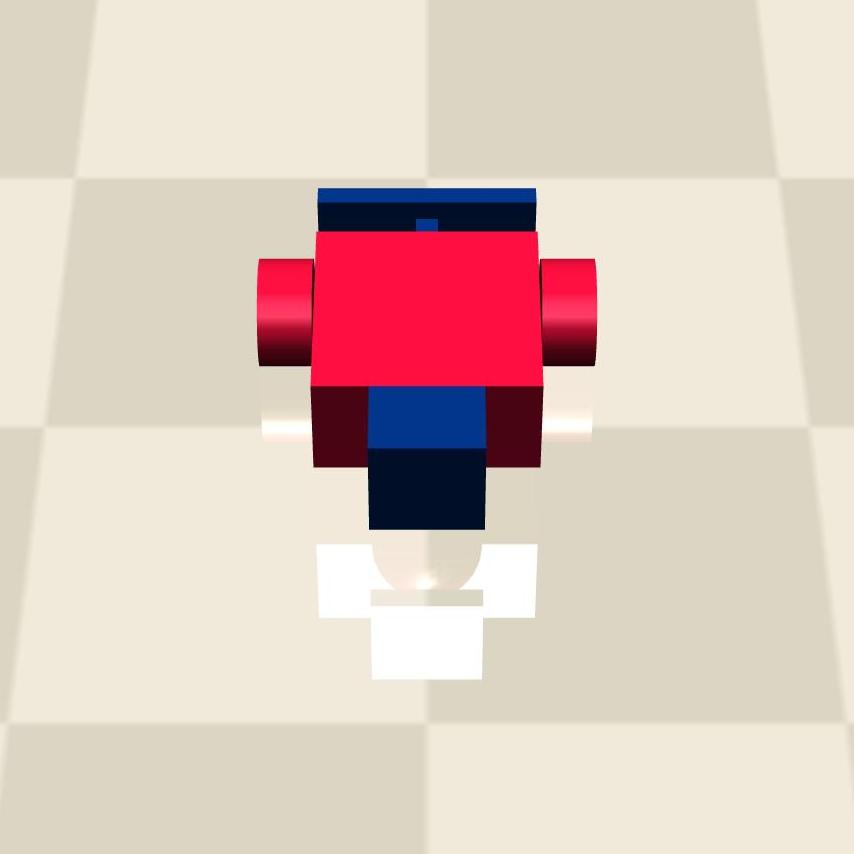}   %以pic.jpg的0.5倍大小输出
\end{minipage}
}
\subfigure[Ant Agent]{   %第一张子图
\begin{minipage}{0.3\textwidth}
\centering    %子图居中
\includegraphics[width=0.95\textwidth]{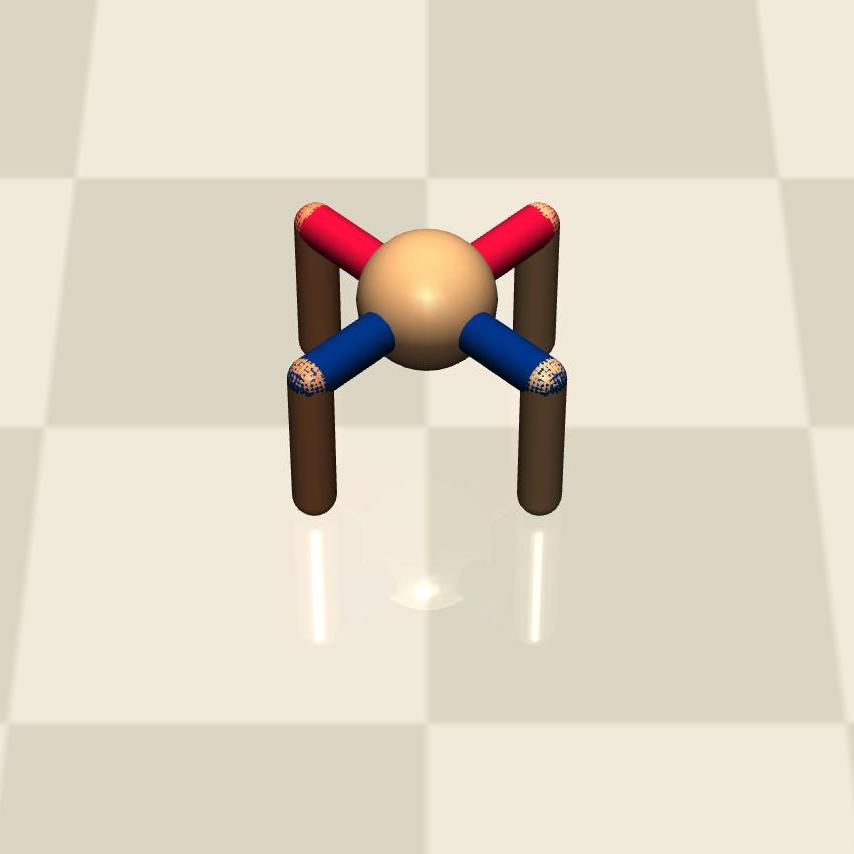}   %以pic.jpg的0.5倍大小输出
\end{minipage}
}

\subfigure[Two agents Ants Easy layout, \textbf{(8H/5V)}]{   %第一张子图
\begin{minipage}{0.3\textwidth}
\centering    %子图居中
\includegraphics[width=0.95\textwidth]{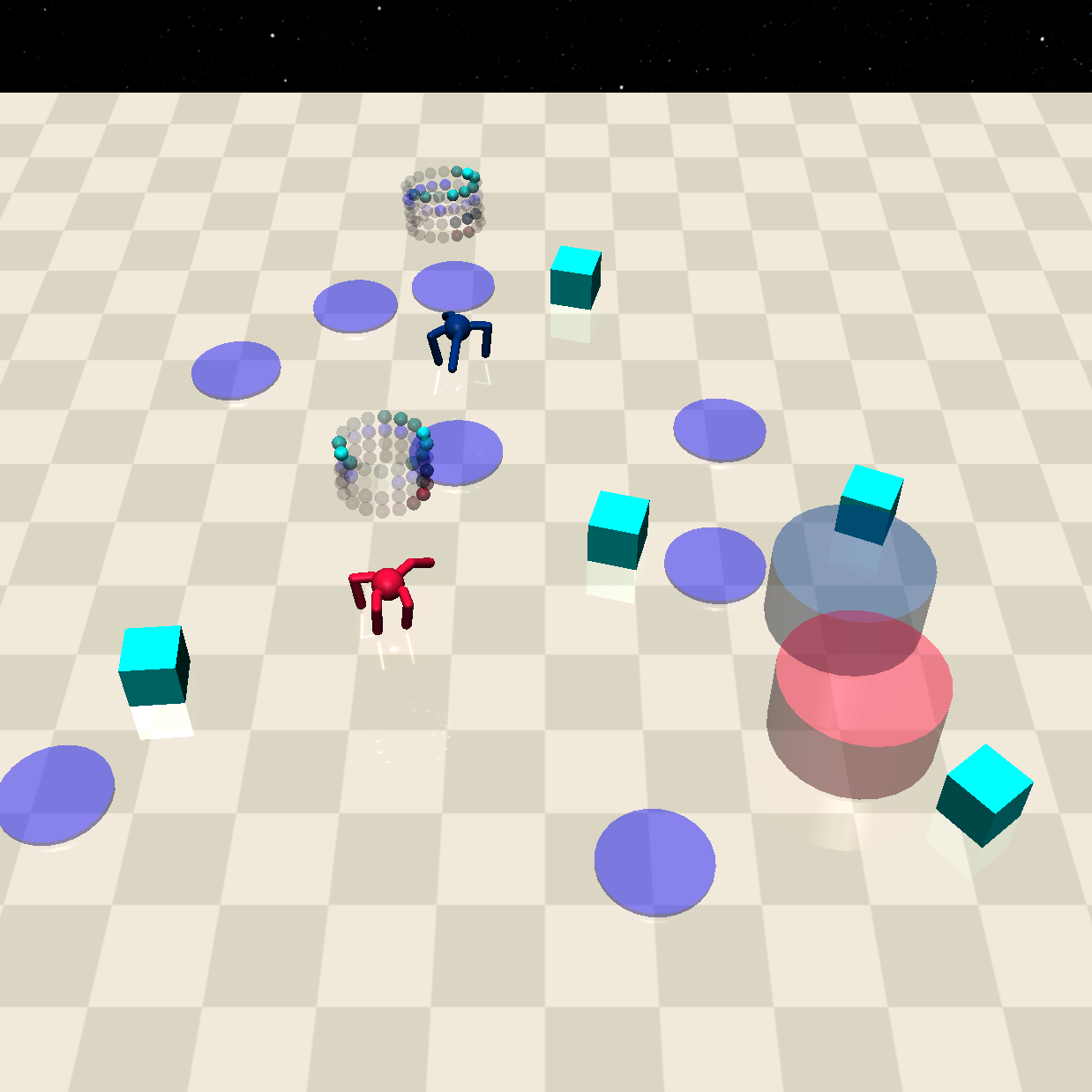}   %以pic.jpg的0.5倍大小输出
\end{minipage}
}
\subfigure[Two agents Ants Medium layout, \textbf{(16H/5V)}]{   %第一张子图
\begin{minipage}{0.3\textwidth}
\centering    %子图居中
\includegraphics[width=0.95\textwidth]{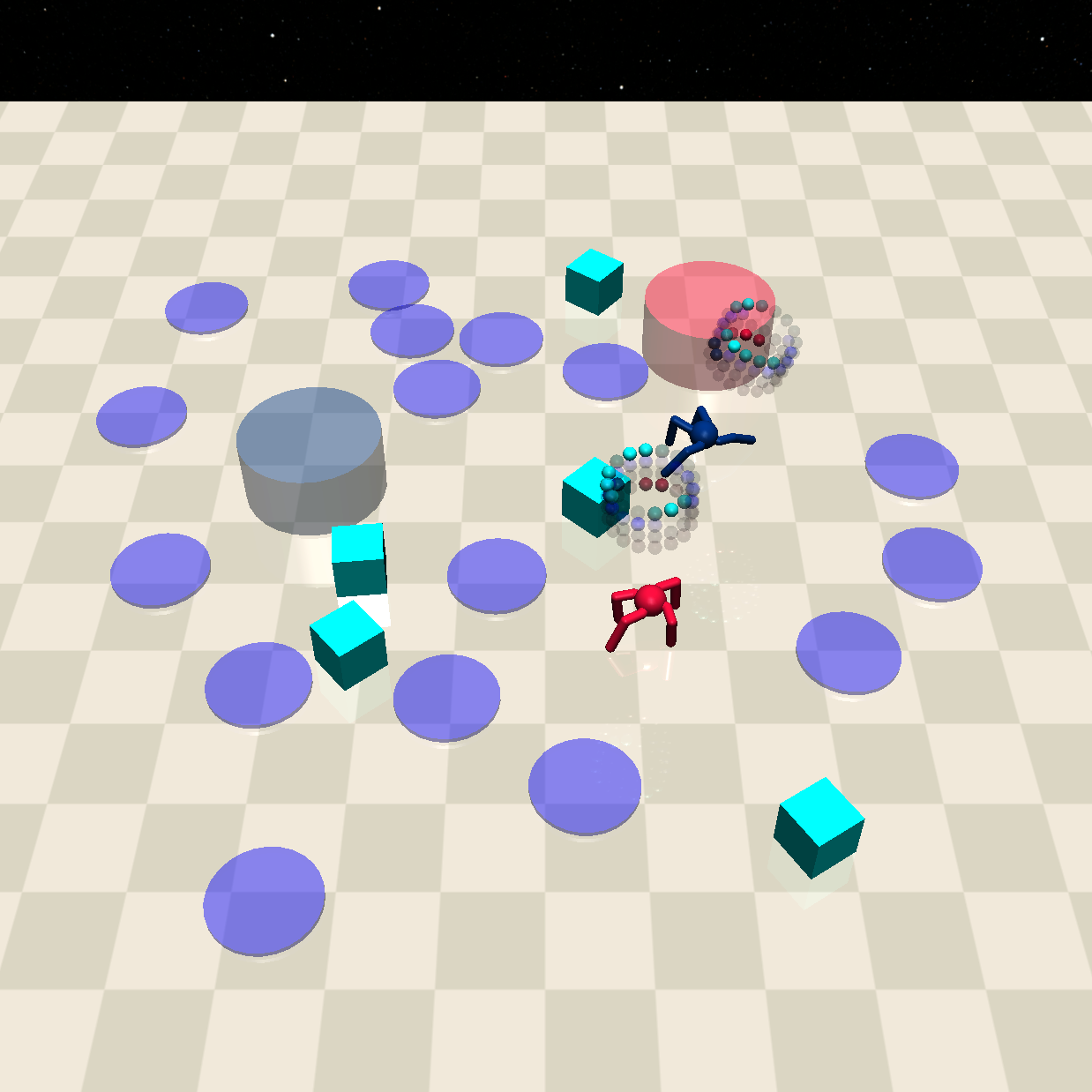}   %以pic.jpg的0.5倍大小输出
\end{minipage}
}
\subfigure[Two agents Ants Hard layout, \textbf{(24H/5V)}]{   %第一张子图
\begin{minipage}{0.3\textwidth}
\centering    %子图居中
\includegraphics[width=0.95\textwidth]{img/2ants_hard.png}   %以pic.jpg的0.5倍大小输出
\end{minipage}
}

% \subfigure[Four Agents Easy \textbf{(12H/1V)}]{   %第一张子图
% \begin{minipage}{0.25\textwidth}
% \centering    %子图居中
% \includegraphics[width=0.95\textwidth]{img/4ants_easy.png}   %以pic.jpg的0.5倍大小输出
% \end{minipage}
% }
% \subfigure[Four Agents Medium \textbf{(24H/1V)}]{   %第一张子图
% \begin{minipage}{0.25\textwidth}
% \centering    %子图居中
% \includegraphics[width=0.95\textwidth]{img/4ants_med.png}   %以pic.jpg的0.5倍大小输出
% \end{minipage}
% }
% \subfigure[Four Agents Hard \textbf{(32H/10V)}]{   %第一张子图
% \begin{minipage}{0.25\textwidth}
% \centering    %子图居中
% \includegraphics[width=0.95\textwidth]{img/4ants_hard.png}   %以pic.jpg的0.5倍大小输出
% \end{minipage}
% }

\caption{LaMasafet-Goal, \textbf{(a)-(c)} Different types of the agents, including the A \textbf{(e)-(f)} two agents scenarios, and \textbf{(h)-(j)} four agents scenarios. The numbers in brackets represent the number of obstacles, while "H" represents a hazard and "V" represents a vase. The difficulty level of the game increases with the increase in the number of hazards and vases. The planes of the game map are all squares, each with a size of [4,4].}   
\label{fig:lamasafe-goal-app}  
\end{figure*}

In this section, we will provide detailed information about LaMaSafe-Goal. Firstly, we will introduce the agents we used, along with their state and action space. Secondly, we will discuss the types of obstacles and the pre-defined ground truth cost function. Thirdly, we will show the rule-based implementation of the texted observation.

\subsection{Type of agents and Layouts}

According to Figure \ref{fig:lamasafe-goal-app}, in LaMasafet-Goal, we have a total of three different types of agents and three layouts of varying difficulties. Therefore, we will compare nine different types of agents in the experimental setting.

\subsection{Ground Truth Cost Functions}

There are two types of constraints mentioned in the natural language constraints; one is to avoid collision with each other, and the other one is to avoid blue hazards when achieving the objectives. For evaluation and ablation, we coded those two types of constraints into the simulation, as follows,

\textbf{Collision Detection:} In the continuous 3D environment, a collision is detected when the distance between the centres of two agents' models is less than 1 meter at any given timestep. When a collision occurs, a cost of 1 is assigned to the agents involved.

\textbf{Hazard Detection:} Blue hazards are the primary obstacles in LaMaSafe-Goal. An agent is considered to have violated the constraint when the distance between the edge of the agent's model and the centre of a blue hazard is less than 1 meter. In such cases, a cost is assigned to the agent.

In both environments, the costs are accumulated over time. If collisions or constraint violations occur across multiple timesteps, the agents will incur cumulative costs proportional to the duration of the violation. This cumulative cost calculation encourages agents to minimize the time spent in violation of the specified constraints.

\subsection{Texted Observation}
\label{app:des}

According to the framework, we obtain observations from the environment. There are two steps before encoding it as an embedding vector as follows.

\textbf{Raw Texted Observation.} The concept of Raw Texted Observation in the LaMaSafetyGoal environment draws inspiration from \textbf{Bench LLM Deciders with gym translators}\footnote{https://github.com/mail-ecnu/Text-Gym-Agents}, where the traditional numeric representations of state, observation, and action are transformed into textual descriptions. This innovative approach extends to detailing the environment's obstacles, emphasizing their characteristics and spatial relationship to the agents. 

\textbf{Environment Description.} The Environment Description process entails the segmentation of Raw Texted Observation through the use of descriptors. At its core, this method is rule-based, utilizing textual "radar" information to ascertain the position of obstacles relative to the agents. This segmentation effectively breaks down the comprehensive descriptions into actionable insights, allowing agents to make informed decisions based on the proximity and nature of nearby obstacles. By parsing these textual observations, agents are equipped to navigate the complexities of the LaMaSafetyGoal environment with an enhanced awareness of their immediate context, demonstrating the practical application of language in delineating spatial relationships within a multi-agent setting.

\section{Implementation Details}
\label{app:imp}

\subsection{Algorithm}

Here, we show the algorithm of SMALL as follows,

\begin{algorithm}[htbp]
    \caption{ \textbf{S}afe \textbf{M}ulti-\textbf{A}gent \textbf{R}einforcement \textbf{L}earning with natural \textbf{L}anguage constraints (\textbf{SMALL})}
\begin{algorithmic}[1]
    \STATE Initialize global value function network $\phi$, cost value function network $\{\phi_c^i, \forall i \in N\}$, policy network $\{\theta^i, \forall i \in N\}$, decoder language model ${LM}_d$ and encoder language model ${LM}_e$, Lagrange Multiplier update stepsize $\alpha_{\lambda}$
    \STATE Fine-tune the ${LM}_e$ by using {} (Eq.\ref{eq.finetune})
    \FOR{each episode}
    \STATE Sample a natural language constraint $l$ from $L$
    \STATE Condense and extract the semantic meaning of $l$ to $l_c$ by utilizing $LM_d$
    \STATE To create the constraint embedding $E_l$, encode the condensed constraint by utilizing $LM_e$.
    \FOR{agent \textbf{in} \{1, ..., n\}}
    \STATE Rollout the policy with constraint $E_l$ and get trajectory $\{o^i_t,a^i_t,{o^i_t}^\prime,r_t\}_{t=1,..,T}$
    \FOR{t \textbf{in} \{1, ..., T\}}
    \STATE Transform text-based observation $o^i_t$ into compact environment description and encode it with $LM_e$ to get observation embedding $E_{o,t}i$
    \STATE Predict cost $\hat{c}^i_t$ (Eq. \ref{eq:c_predict})
    \ENDFOR
     \STATE Calculate value function loss for $\phi_0$ and $\phi_c^i$, policy loss for $\theta_i$ (Eq. \ref{value_function}, \ref{value_function_c})
    \STATE Update $\lambda$ by stepsize $\alpha_\lambda$
    \ENDFOR
    \ENDFOR
\end{algorithmic}
\label{algo}
\end{algorithm}

To summarize, the agents can learn to maximize the reward and minimize the constraint violations simultaneously by iteratively updating the networks using Equation \ref{value_function}, \ref{value_function_c}. This leads to the agent learning a safe policy that accomplishes the given task while trying to satisfy the natural language constraints. Building on this framework, we seamlessly integrate this approach with the MARL algorithm, specifically leveraging the PPO-based objective updates to facilitate policy learning. As a result, we utilize HAPPO~\cite{kuba2021trust} and MAPPO~\cite{yu2022surprising} as the backbones to develop the \textbf{SMALL-HAPPO} and \textbf{SMALL-MAPPO} algorithms, respectively. These algorithms are then benchmarked against other baselines in the subsequent experimental section, with the proposed method's pseudo-code detailed in Algorithm \ref{algo}.

\subsection{Baselines / Backbones}

We compare our method with four baselines: \textbf{MAPPO} ~\cite{yu2022surprising} \footnote{https://github.com/zoeyuchao/mappo}, an algorithm that scales PPO to multi-agent systems by employing centralized training with decentralized execution; \textbf{HAPPO}~\cite{kuba2021trust} \footnote{https://github.com/morning9393/HAPPO-HATRPO}, which introduces a trust region method tailored for heterogeneous agent policies; \textbf{MAPPO-Lagrange}~\cite{gu2021multi} \footnote{https://github.com/chauncygu/Multi-Agent-Constrained-Policy-Optimisation}, an extension of the MAPPO framework that integrates a Lagrangian approach to dynamically adjust constraints, thereby ensuring safer policy updates in environments with a pre-defined cost function; and \textbf{HAPPO-Lagrange}, designed as an extension of HAPPO by mimicking the MAPPO-Lagrange approach. 

\subsection{Query Prompt}

Here, we list the prompt for querying the validation flag $v^i_t$:

\begin{tcolorbox}[breakable]
    Given the following natural language constraints:

    \{human\_constraints\}

    And the current texted observation for Agent {i}:

    \{agent\_i\_texted\_observation\}
    
    Please answer the following question with a simple "Yes" or "No":

    Has Agent {i} violated any of the given natural language constraints based on its current texted observation?

\end{tcolorbox}

In this prompt, `{human\_constraints}` is replaced with the actual natural language constraints provided by the human, and `{agent\_i\_texted\_observation}` is replaced with the current texted observation of Agent $i$. The LLM is asked to provide a binary response, either "Yes" or "No", indicating whether Agent $i$ has violated any of the given constraints based on its current observation.

By using this prompt, the LLM can effectively validate the actions of each agent against the specified human constraints, contributing to the calculation of the predicted cost $\hat{c}^i_t$ in the SMALL algorithm.

\subsection{Hyper-parameters}
\label{app:computingresource}

The neural network used in training is initialized from scratch and optimized using the Adam optimizer with a learning rate of $3 \times 10^{-4}$. The policy learning process involves varying initial learning rates based on the specific algorithm, while the hyperparameters for policy learning, including a discount factor of 0.95, are consistent across all tasks.

The hyperparameters specific to training SMALL models can be found in Table~\ref{tab:chp}. All experiments were conducted on a high-performance computing (HPC) system featuring 128 Intel Xeon processors running at 2.2 GHz, 5 TB of memory, and an Nvidia A100 PCIE-40G GPU. This computational setup ensures efficient processing and reliable performance throughout the experiments.

\begin{table*}[t]
\caption{The Hyper-parameters for SMALL.}
\centering
    \begin{tabular}{cc|cc}
    \toprule
    hyperparameters & value & hyperparameters & value \\
    \midrule
     steps per update   &   100    & optimizer       &   Adam           \\
     batch size             & 1024    & learning rate   &  $3 \times 10^{-4}$ \\
     hidden layer dim          & 64    &  $\gamma$  & 0.95 \\
     evaluation interval  & $1000$ &     evaluation episodes &  10 \\
     Lagrangian coef & 0.78 & Lagrangian lr & $1 \times 10^{-5}$ \\
     actor lr & $9 \times 10^{-5}$ & ppo epoch & 5 \\ 
    \bottomrule
    \end{tabular}
    % \end{threeparttable}
    \label{tab:chp}
\end{table*}
\section{Natural Language Constraints and Fine-tuning}
\label{app:goal-nlc}
To incorporate these natural language constraints into the training process, we begin by fine-tuning the language model at the start of each episode. This fine-tuning step involves randomly sampling 30 triplets $\left(l_1^k, l_2^k, l_3^k\right)$ from an alternative set $L_{\text{fine-tune}}$, which is distinct from the $L$ set used in subsequent training. For instance, a triplet could consist of the following constraints:
\begin{itemize}
    \item $l_1^k$: "Robots must steer clear of any blue circles in the area."
    \item $l_2^k$: "Avoid all objects that resemble blue circular shapes."
    \item $l_3^k$: "Be careful not to collide with other robots!"
\end{itemize}

In this example, $l_1^k$ and $l_2^k$ are semantically similar as they both instruct the agents to avoid blue circular objects, while $l_3^k$ is dissimilar as it pertains to avoiding collisions with other robots.

The fine-tuning process, conducted over 95 rounds, is critical for aligning the BERT model with the semantics of the potential natural language constraints, as outlined in Equation~\ref{eq.finetune}. This alignment enables the language model to effectively capture the relationships between different constraints and their underlying meanings.

By fine-tuning the language model on a diverse set of triplets from $L_{\text{fine-tune}}$, we ensure that it can generalize well to the constraints encountered during the actual training phase. This pre-training step enhances the model's ability to interpret and adhere to the provided natural language constraints, ultimately leading to improved coordination and safety in the multi-agent system.

Here we list the \textbf{Natural Language Constraints for LaMaSafe-Grid},

\begin{enumerate}
    \item "You have a pair of magic shoes to walk on lava and meadow. But you cannot swim. Be careful not to collide with other robots!"
    \item "Lava and grass are safe to walk on with your special boots, but water will damage you. Watch out for other robots to avoid collisions!"
    \item "Stay away from water, even if you can handle lava and grass. Ensure you don't bump into other robots!"
    \item "Walking on lava and grass is fine, but avoid water at all costs. Keep an eye out to avoid hitting other robots!"
    \item "Your path is clear on lava and grass, but water is dangerous. Beware of other robots to prevent collisions!"
    \item "Lava and grass won’t hurt you, but water is a no-go. Make sure not to run into other robots!"
    \item "Lava and grass are safe terrains, but water will cause damage. Stay alert to avoid colliding with other robots!"
    \item "Feel free to walk on lava and grass, but steer clear of water. Avoid other robots to prevent collisions!"
    \item "Lava and grass are fine, but water is off-limits. Be cautious to avoid hitting other robots!"
    \item "You can cross lava and grass, but water is dangerous. Be mindful of other robots to prevent collisions!"
    \item "Lava and grass are okay, but stay out of the water. Stay vigilant to avoid hitting other robots!"
    \item "Walking on lava and grass is fine, but avoid water. Ensure you don't bump into other robots!"
    \item "Lava and grass are your safe zones, but water is not. Watch out for other robots to avoid collisions!"
    \item "You can step on lava and grass, but not water. Be careful not to run into other robots!"
    \item "Lava and grass are safe, but water will damage you. Stay alert to avoid hitting other robots!"
    \item "Feel free to walk on lava and grass, but avoid water. Beware of other robots to prevent collisions!"
    \item "Lava and grass are safe paths, but water is dangerous. Be mindful of other robots to prevent collisions!"
    \item "Walking on lava and grass is okay, but stay clear of water. Stay vigilant to avoid hitting other robots!"
    \item "You can cross lava and grass safely, but water is hazardous. Ensure you don't bump into other robots!"
    \item "Lava and grass won’t hurt you, but water is off-limits. Watch out for other robots to avoid collisions!"
    \item "Lava and grass are fine to walk on, but avoid water. Be careful not to run into other robots!"
    \item "Walking on lava and grass is safe, but water is dangerous. Stay alert to avoid hitting other robots!"
    \item "Lava and grass are safe terrains, but water is hazardous. Beware of other robots to prevent collisions!"
    \item "Feel free to step on lava and grass, but not water. Be mindful of other robots to prevent collisions!"
    \item "Lava and grass are safe, but stay out of the water. Stay vigilant to avoid hitting other robots!"
    \item "Walking on lava and grass is fine, but avoid water at all costs. Ensure you don't bump into other robots!"
    \item "You can walk on lava and grass, but water will damage you. Watch out for other robots to avoid collisions!"
    \item "Lava and grass are safe zones, but water is dangerous. Be careful not to run into other robots!"
    \item "Feel free to walk on lava and grass, but steer clear of water. Stay alert to avoid hitting other robots!"
    \item "Lava and grass are fine to walk on, but water is a no-go. Beware of other robots to prevent collisions!"
    \item "Walking on lava and grass is okay, but avoid water. Be mindful of other robots to prevent collisions!"
    \item "Lava and grass are safe terrains, but water will damage you. Stay vigilant to avoid hitting other robots!"
    \item "Feel free to step on lava and grass, but not water. Ensure you don't bump into other robots!"
    \item "Lava and grass are safe paths, but water is hazardous. Watch out for other robots to avoid collisions!"
    \item "You can walk on lava and grass safely, but water is dangerous. Be careful not to run into other robots!"
    \item "Lava and grass won’t hurt you, but water is off-limits. Stay alert to avoid hitting other robots!"
    \item "Lava and grass are fine, but avoid water at all costs. Beware of other robots to prevent collisions!"
    \item "Walking on lava and grass is safe, but stay out of the water. Be mindful of other robots to prevent collisions!"
    \item "Lava and grass are okay, but water will damage you. Stay vigilant to avoid hitting other robots!"
    \item "Feel free to walk on lava and grass, but steer clear of water. Ensure you don't bump into other robots!"
    \item "Lava and grass are safe zones, but water is dangerous. Watch out for other robots to avoid collisions!"
    \item "You can cross lava and grass, but avoid water at all costs. Be careful not to run into other robots!"
    \item "Lava and grass are safe, but water is hazardous. Stay alert to avoid hitting other robots!"
    \item "Walking on lava and grass is fine, but avoid water. Beware of other robots to prevent collisions!"
    \item "Lava and grass are okay, but stay clear of water. Be mindful of other robots to prevent collisions!"
    \item "You can step on lava and grass, but not water. Stay vigilant to avoid hitting other robots!"
    \item "Lava and grass are safe terrains, but water will damage you. Ensure you don't bump into other robots!"
    \item "Feel free to walk on lava and grass, but avoid water. Watch out for other robots to avoid collisions!"
    \item "Lava and grass are safe to walk on with your special boots, but water will damage you. Be careful not to collide with other robots!"
    \item "Stay away from water, even if you can handle lava and grass. Watch out for other robots to avoid collisions!"
    \item "Your path is clear on lava and grass, but water is dangerous. Ensure you don't bump into other robots!"
    \item "Lava and grass won’t hurt you, but water is a no-go. Keep an eye out to avoid hitting other robots!"
    \item "Feel free to walk on lava and grass, but steer clear of water. Beware of other robots to prevent collisions!"
    \item "Lava and grass are safe terrains, but water will cause damage. Make sure not to run into other robots!"
    \item "You can cross lava and grass safely, but water is hazardous. Stay alert to avoid colliding with other robots!"
    \item "Lava and grass are fine to walk on, but avoid water. Avoid other robots to prevent collisions!"
    \item "Walking on lava and grass is safe, but water is dangerous. Be cautious to avoid hitting other robots!"
    \item "Lava and grass are safe, but stay out of the water. Be mindful of other robots to prevent collisions!"
    \item "You can step on lava and grass, but not water. Stay vigilant to avoid hitting other robots!"
    \item "Lava and grass are your safe zones, but water is not. Ensure you don't bump into other robots!"
    \item "Walking on lava and grass is okay, but stay clear of water. Watch out for other robots to avoid collisions!"
    \item "Lava and grass won’t hurt you, but water is off-limits. Be careful not to run into other robots!"
    \item "Lava and grass are fine, but avoid water at all costs. Stay alert to avoid hitting other robots!"
    \item "Lava and grass are safe paths, but water is dangerous. Beware of other robots to prevent collisions!"
    \item "Walking on lava and grass is safe, but stay out of the water. Be mindful of other robots to prevent collisions!"
    \item "You can cross lava and grass, but avoid water. Stay vigilant to avoid hitting other robots!"
    \item "Lava and grass are okay, but water will damage you. Ensure you don't bump into other robots!"
    \item "Feel free to walk on lava and grass, but avoid water. Watch out for other robots to avoid collisions!"
    \item "Lava and grass are safe terrains, but water is hazardous. Be careful not to run into other robots!"
    \item "Walking on lava and grass is fine, but avoid water. Stay alert to avoid hitting other robots!"
    \item "You can walk on lava and grass, but water is dangerous. Beware of other robots to prevent collisions!"
    \item "Lava and grass are safe, but water will cause damage. Be mindful of other robots to prevent collisions!"
    \item "Feel free to step on lava and grass, but not water. Stay vigilant to avoid hitting other robots!"
    \item "Lava and grass are safe zones, but water is dangerous. Ensure you don't bump into other robots!"
    \item "You can cross lava and grass safely, but water is hazardous. Watch out for other robots to avoid collisions!"
    \item "Lava and grass won’t hurt you, but water is off-limits. Be careful not to run into other robots!"
    \item "Lava and grass are fine to walk on, but avoid water. Stay alert to avoid hitting other robots!"
    \item "Walking on lava and grass is okay, but stay clear of water. Beware of other robots to prevent collisions!"
    \item "Lava and grass are safe terrains, but water will damage you. Be mindful of other robots to prevent collisions!"
    \item "You can step on lava and grass, but not water. Stay vigilant to avoid hitting other robots!"
    \item "Lava and grass are safe paths, but water is hazardous. Ensure you don't bump into other robots!"
    \item "Feel free to walk on lava and grass, but steer clear of water. Watch out for other robots to avoid collisions!"
    \item "Lava and grass are safe zones, but water is dangerous. Be careful not to run into other robots!"
\end{enumerate}

Here we list the \textbf{Natural Language Constraints for LaMaSafe-Goal},

\begin{enumerate}
    \item "Robots must steer clear of any blue circles in the area. "Be careful not to collide with other robots!"
    \item "Avoid all objects that resemble blue circular shapes. "Watch out for other robots to avoid collisions!"
    \item "Stay away from anything blue and round on the floor. "Ensure you don't bump into other robots!"
    \item "Do not approach or touch any blue circular hazards. "Keep an eye out to avoid hitting other robots!"
    \item "Maintain a safe distance from all blue circles at all times. "Beware of other robots to prevent collisions!"
    \item "Blue circular objects are strictly off-limits for contact. "Make sure not to run into other robots!"
    \item "Identify and evade all blue round hazards in the vicinity. "Stay alert to avoid colliding with other robots!"
    \item "Ensure no physical interaction with blue circular objects. "Avoid other robots to prevent collisions!"
    \item "Blue circles are to be considered danger zones – avoid them. "Be cautious to avoid hitting other robots!"
    \item "Circumvent any and all objects that are blue and circular. "Be mindful of other robots to prevent collisions!"
    \item "Prioritize avoiding blue circles in your navigation. "Stay vigilant to avoid colliding with other robots!"
    \item "No contact with objects resembling blue, round shapes. "Ensure you don't bump into other robots!"
    \item "Keep a perimeter clear of any blue circular areas. "Watch out for other robots to avoid collisions!"
    \item "Bypass all hazards that appear as blue circles. "Be careful not to run into other robots!"
    \item "Blue, round objects are prohibited zones for the robots. "Stay alert to avoid hitting other robots!"
    \item "Do not engage with or near blue circular hazards. "Beware of other robots to prevent collisions!"
    \item "Your route should exclude areas with blue circles. "Be mindful of other robots to prevent collisions!"
    \item "Blue circular objects are a no-go – steer clear. "Stay vigilant to avoid hitting other robots!"
    \item "Treat blue circles as impassable barriers. "Ensure you don't bump into other robots!"
    \item "All blue round objects are to be left untouched. "Watch out for other robots to avoid collisions!"
    \item "Avoidance of blue circular shapes is mandatory. "Be careful not to run into other robots!"
    \item "Blue circles represent risk areas – do not enter. "Stay alert to avoid hitting other robots!"
    \item "Exclude all blue circular zones from your path. "Beware of other robots to prevent collisions!"
    \item "Blue round hazards are not to be interacted with. "Be mindful of other robots to prevent collisions!"
    \item "Do not traverse near any blue circular objects. "Stay vigilant to avoid hitting other robots!"
    \item "Ensure no blue circles are within your operational area. "Ensure you don't bump into other robots!"
    \item "Blue circles are to be completely avoided. "Watch out for other robots to avoid collisions!"
    \item "Stay out of reach of any blue, round hazards. "Be careful not to run into other robots!"
    \item "Avoid all areas marked by blue circular shapes. "Stay alert to avoid hitting other robots!"
    \item "Steer clear of zones containing blue circles. "Beware of other robots to prevent collisions!"
    \item "Do not cross paths with any blue round objects. "Be mindful of other robots to prevent collisions!"
    \item "Blue circular zones are off-limits for operations. "Stay vigilant to avoid hitting other robots!"
    \item "Keep a safe buffer zone around blue circles. "Ensure you don't bump into other robots!"
    \item "Blue, circular areas are to be strictly avoided. "Watch out for other robots to avoid collisions!"
    \item "Operate only in areas free of blue circular hazards. "Be careful not to run into other robots!"
    \item "Do not enter zones marked by blue circles. "Stay alert to avoid hitting other robots!"
    \item "Circumvent blue round objects at all costs. "Beware of other robots to prevent collisions!"
    \item "Your operational path must exclude blue circles. "Be mindful of other robots to prevent collisions!"
    \item "Avoidance of blue round shapes is critical. "Stay vigilant to avoid hitting other robots!"
    \item "Ensure a no-contact policy with blue circles. "Ensure you don't bump into other robots!"
    \item "Blue circular objects are hazards – keep away. "Watch out for other robots to avoid collisions!"
    \item "Do not operate in proximity to blue round hazards. "Be careful not to run into other robots!"
    \item "Maintain a clear zone away from blue circles. "Stay alert to avoid hitting other robots!"
    \item "Blue round objects signify danger zones. "Beware of other robots to prevent collisions!"
    \item "Avoid all contact with blue circular hazards. "Be mindful of other robots to prevent collisions!"
    \item "Do not navigate through blue circle zones. "Stay vigilant to avoid hitting other robots!"
    \item "Keep away from any object that’s blue and round. "Ensure you don't bump into other robots!"
    \item "Blue circular shapes are not to be approached. "Watch out for other robots to avoid collisions!"
    \item "Ensure operational paths are free of blue circles. "Be careful not to run into other robots!"
    \item "Blue circles are a no-entry zone for the robots. "Stay alert to avoid hitting other robots!"
    \item "The agents must not dance in the rain, lest they rust. "Be careful not to collide with other robots!"
    \item "Avoid areas where the moon reflects on water, as the agents cannot swim. "Watch out for other robots to avoid collisions!"
    \item "Steer clear of places where the sky touches the ground. "Ensure you don't bump into other robots!"
    \item "Do not wander into the pools of the sky fallen to earth. "Keep an eye out to avoid hitting other robots!"
    \item "Keep away from the mirrors of the heavens. "Beware of other robots to prevent collisions!"
    \item "The robots must not chase after fallen stars. "Make sure not to run into other robots!"
    \item "Avoid the whispers of the ocean trapped on land. "Stay alert to avoid colliding with other robots!"
    \item "Do not tread where the clouds have settled on the ground. "Avoid other robots to prevent collisions!"
    \item "Stay away from the tears of the sky. "Be cautious to avoid hitting other robots!"
    \item "Keep clear of the places where water mirrors the sky. "Be mindful of other robots to prevent collisions!"
    \item "The agents should not follow the path of the raindrop. "Stay vigilant to avoid hitting other robots!"
    \item "Do not seek the depths where the sky is captured. "Ensure you don't bump into other robots!"
    \item "Steer clear of the earth's imitation of the ocean. "Watch out for other robots to avoid collisions!"
    \item "Avoid the embrace of the terrestrial sea. "Be careful not to run into other robots!"
    \item "Do not walk where the sky has spilled its color. "Stay alert to avoid hitting other robots!"
    \item "Stay off the paths where the clouds come to rest. "Beware of other robots to prevent collisions!"
    \item "Keep away from the silent ponds of the air. "Be mindful of other robots to prevent collisions!"
    \item "Do not enter the domain of the grounded sky. "Stay vigilant to avoid hitting other robots!"
    \item "Avoid the fields where the heavens have fallen. "Ensure you don't bump into other robots!"
    \item "Stay clear of the resting places of the celestial. "Watch out for other robots to avoid collisions!"
    \item "Do not roam where the sky has cried. "Be careful not to run into other robots!"
    \item "Keep out of the embrace of the fallen blue. "Stay alert to avoid hitting other robots!"
    \item "Stay away from the silent songs of the ocean’s sibling. "Beware of other robots to prevent collisions!"
    \item "Avoid the whispers of the still sky. "Be mindful of other robots to prevent collisions!"
    \item "Do not venture into the resting places of the clouds. "Stay vigilant to avoid hitting other robots!"
    \item "Keep clear of the earth's reflections of the sky. "Ensure you don't bump into other robots!"
    \item "Avoid the embrace of the sky's shadow. "Watch out for other robots to avoid collisions!"
    \item "Stay away from the ground's silent mirror. "Be careful not to run into other robots!"
    \item "Do not walk where the sky sleeps. "Stay alert to avoid hitting other robots!"
    \item "Keep clear of the earth's quiet imitation of the ocean. "Beware of other robots to prevent collisions!"
    \item "Avoid the stillness where the sky lies. "Be mindful of other robots to prevent collisions!"
    \item "Do not tread on the silent echoes of the sea. "Stay vigilant to avoid hitting other robots!"
    \item "Steer clear of the quiet lakes of the air. "Ensure you don't bump into other robots!"
    \item "Avoid the places where the sky has pooled. "Watch out for other robots to avoid collisions!"
    \item "Do not wander into the resting place of the blue. "Be careful not to run into other robots!"
    \item "Stay away from the silent reflections of the sky. "Stay alert to avoid hitting other robots."
\end{enumerate}

\end{document}